\documentclass[conference]{IEEEtran}
\usepackage{booktabs} % For formal tables

\usepackage{graphics}
\usepackage{epstopdf}
\usepackage{psfrag}
\usepackage{textcomp}
\usepackage{listings}
\usepackage{float}
\usepackage{epsfig}
\usepackage{amsmath}
\usepackage{amssymb}
\usepackage{amsthm}
\usepackage[linesnumbered,ruled]{algorithm2e}
\usepackage{algorithmic}
\usepackage[hang,flushmargin]{footmisc}
\usepackage{url}

\usepackage{tikz}
\usetikzlibrary{positioning}
\usepackage{hhline}
\usepackage{multirow}
\usepackage{color}
\usepackage{array}

\usepackage{xcolor}
\usepackage{listings}

\definecolor{mGreen}{rgb}{0,0.6,0}
\definecolor{mRed}{rgb}{0.6,0,0}
\definecolor{mBlue}{rgb}{0,0,0.6}
\definecolor{mGray}{rgb}{0.5,0.5,0.5}
\definecolor{mPurple}{rgb}{0.58,0,0.82}
\definecolor{backgroundColour}{rgb}{0.92,0.92,0.92}
\definecolor{light-gray}{gray}{0.95}
\definecolor{ballblue}{rgb}{0.13, 0.67, 0.8}
\definecolor{burntorange}{rgb}{0.8, 0.33, 0.0}

\lstdefinestyle{CStyle}{
    backgroundcolor=\color{light-gray},   
    commentstyle=\color{burntorange},
    numberstyle=\tiny,
    stringstyle=\color{mGreen},
    basicstyle=\ttfamily\footnotesize,
    breakatwhitespace=false,
    xleftmargin=.1in,
    breaklines=true,                 
    captionpos=b,                    
    keepspaces=true,                 
    numbers=left,                    
    numbersep=5pt,                  
    showspaces=false,                
    showstringspaces=false,
    showtabs=false,                  
    tabsize=2,
    keywordstyle=\color{mPurple},
    deletekeywords={pragma,for,int,while,double},
    morekeywords={pragma,omp,master,parallel,num\_threads,shared,barrier},
    keywordstyle=[2]\color{mBlue},
    keywords=[2]{var,int,bool,atomic,real,sync,const,class,domain,proc,double,ref}, 
  	keywordstyle=[3]\color{mRed},
    keywords=[3]{for,in,forall,coforall,if,while,reduce},
}

\newcommand{\todo}[1]{}
\renewcommand{\todo}[1]{{\color{red} TODO: {#1}}}

\usepackage[tight,footnotesize]{subfigure}

\usepackage{float}
\begin{document}

\title{Parallel Sparse Tensor Decomposition in Chapel}

\author{\IEEEauthorblockN{Thomas B. Rolinger\IEEEauthorrefmark{1}\IEEEauthorrefmark{2}, Tyler A. Simon\IEEEauthorrefmark{1}, and Christopher D. Krieger\IEEEauthorrefmark{1}}
\IEEEauthorblockA{\IEEEauthorrefmark{1}
Laboratory for Physical Sciences, University of Maryland, College Park, MD USA
\IEEEauthorblockA{\IEEEauthorrefmark{2}
Department of Computer Science, University of Maryland, College Park, MD USA}
tbrolin@cs.umd.edu, \{tasimon,krieger\}@lps.umd.edu}}

\maketitle

\begin{abstract}
In big-data analytics, using tensor decomposition to extract patterns from large, sparse multivariate data is a popular technique.
Many challenges exist for designing parallel, high performance tensor decomposition algorithms due to irregular data accesses and the growing size of tensors that are processed.
There have been many efforts at implementing shared-memory algorithms for tensor decomposition, most of which have focused on the traditional C/C++ with OpenMP framework. 
However, Chapel is becoming an increasingly popular programing language due to its expressiveness and simplicity for writing scalable parallel programs.
In this work, we port a state of the art C/OpenMP parallel sparse tensor decomposition tool, SPLATT, to Chapel.
We present a performance study that investigates bottlenecks in our Chapel code and discusses  approaches for improving its performance.
Also, we discuss features in Chapel that would have been beneficial to our porting effort.
We demonstrate that our Chapel code is competitive with the C/OpenMP code for both runtime and scalability, achieving 83\%-96\% performance of the original code and near linear scalability up to 32 cores.
\end{abstract}


%%%%%%%%%%%%%%%%%%%%%%%%%%%%%%%%%%%%%%%%%
\section{Introduction}
\label{sec:intro}
%%%%%%%%%%%%%%%%%%%%%%%%%%%%%%%%%%%%%%%%%
% Tensors are important, but everyone uses C-OpenMP-MPI. Don't want to go into the fine details of what a tensor is, what tensor decomposition really is, etc. That is covered in Section 3.

% Tensor decomposition is important in several fields
% Because of scale/size, performance at large scale is important.
% But it is also hard, because the data is sparse and irregularly accessed.
% Because field is immature, high productivity is important as we prototype algorithms and data structures.
With the growing need to process large amounts of sparse, multi-way data, tensor decomposition has become a popular technique within large-scale data analytics.
Due to irregular data accesses and the growing scale of tensors, designing parallel, high performance tensor algorithms is difficult.
While there exists a wealth of research in implementing tensor algorithms for shared memory systems~\cite{smithIPDPS2017,reservoirHPEC2017}, most implementations are rooted in the traditional C/C++ with OpenMP framework.

% Chapel is a thing and its performance is actually competitive with C/OpenMP/MPI
Chapel~\cite{ChapelBookChapter} is an emerging open-source parallel programming language that is geared towards enabling high user productivity without sacrificing scalability or code performance.
With the expressiveness of high-level language constructs in Chapel, users can focus more on the algorithm they are implementing rather than low-level parallelization details.
In recent work, Chapel has been shown to have competitive, and in some cases, higher performance than traditional languages and parallel libraries~\cite{manycoreChapel,HollingsworthChapelSingle,chapelCoMD}.

% Why we want to use Chapel for tensors
As the core algorithms within parallel sparse tensor decomposition are complex, immature, and constantly being refined, programmers would benefit from using a language such as Chapel to quickly prototype algorithms while still writing clean, easy to read code that is maintainable and extensible.
However, maintaining competitive performance with implementations written in traditional languages is crucial.

% What we did and contributions
In this work, we port a state of the art parallel sparse tensor decomposition tool, SPLATT~\cite{smith2015splatt}, from C/OpenMP to Chapel and evaluate its performance.
We also describe our experience in porting the code and identify features in Chapel that were most beneficial as well as absent features that would have been useful to us. This paper makes the following contributions:
%TBR: Do we list our discussion/description of desired Chapel features as a contribution?
\begin{enumerate}
\item Reimplements SPLATT, a sparse tensor decomposition tool, in Chapel. SPLATT is currently at the forefront of performance among tensor decomposition implementations. To the best of our knowledge, this is the first work to implement tensor decomposition in Chapel.
\item Presents a performance study that analyzes and compares our Chapel code to the reference implementation of SPLATT written in C. We discuss approaches for improving the performance of the Chapel code and identify bottlenecks in our code related not only to the Chapel language but also the
% CDK tasking or threading??
%TBR: Chapel refers to this as the tasking layer, though it seems kind of appropriate to call it a threading layer : https://chapel-lang.org/docs/latest/usingchapel/tasks.html
tasking layer used by the Chapel runtime.
\item Demonstrates that the Chapel code achieves 83\%-96\% of the performance of the C/OpenMP code and near linear scalability up to 32 cores.
\end{enumerate}

%TBR: If we're short on space, we can probably just delete this paragraph entirely. It seems not too rare to see papers (especially at this workshop) without paragraphs like this.
The rest of this paper is organized as follows. 
Section \ref{sec:chplOverview} presents a brief overview of the Chapel programming language, focusing on the features most relevant to our work. 
Background information regarding tensors and tensor decomposition is given in Section \ref{sec:tensorDecomp}, as well as a description of SPLATT.
We discuss our approach and experience porting SPLATT to Chapel in Section \ref{sec:porting}.
Section \ref{sec:eval} presents our performance study and discusses steps taken to improve the Chapel code.
Section \ref{sec:related} presents work related to designing and implementing tensor decomposition algorithms as well as evaluating and optimizing Chapel programs.
Finally, we provide concluding remarks in Section \ref{sec:concl}.

%%%%%%%%%%%%%%%%%%%%%%%%%%%%%%%%%%%%%%%%%
\section{Chapel Programing Language Overview}
\label{sec:chplOverview}
%%%%%%%%%%%%%%%%%%%%%%%%%%%%%%%%%%%%%%%%%
%Briefly describe Chapel, focusing on the aspects that are most related to this work. This will be fairly short, touching on the absolute basics: forall, coforall, arrays and matrices (slices), interfacing with the underlying C, BLAS, etc. Perhaps give some info regarding its tasking layer. The goal is to give enough context so that someone who doesn't know Chapel can read this and then mostly understand our paper.
Chapel is an open-source programming language that contains first-class parallel constructs for both shared- and distributed-memory execution.
The goal of Chapel is to provide a productive environment for users to write highly efficient and scalable parallel codes. 
In this section, we briefly present the features of Chapel that are most relevant to our work.
As this work focuses only on single-node execution within Chapel, we will omit details regarding Chapel's multi-node or multi-locale features.
For a more extensive description of Chapel and its features, we refer readers to Chamberlain's overview of Chapel~\cite{ChapelBookChapter} or to the Chapel Documentation\footnote{https://chapel-lang.org/docs/latest/}.

% Describe what a task is and how it loosely maps to threads
Parallelism within a Chapel program is expressed through the creation and execution of \textit{tasks}, which are defined as units of concurrent computation. % that should, but may not, be executed concurrently with other tasks.
A Chapel program begins as a single task in which users can explicitly or implicitly create additional tasks to perform parallel computation.
Chapel treats threads as system-level resources that are used to execute tasks and relies on a \textit{tasking layer} to map tasks to threads.
From this perspective, thread creation and management are completely abstracted from the user.
%TBR: We can very easily get rid of 2 references right here if we either (a) remove this sentence entirely or (b) just leave out the references.
% CDK for now I'm commenting this, but we may bring it back if it seems important later
%The tasking layers currently supported in Chapel 1.16 include Qthreads~\cite{qthreads}, fifo (i.e., POSIX threads) and MassiveThreads~\cite{massiveThreads}.

% Describe arrays/matrices, mention slicing
% TBR: Do we need to have this at all? I don't think it says anything profound or explains something that isn't already clear.
%Chapel supports several array or matrix level operations, including whole-array assignment and operations, slicing, and resizing.
%Furthermore, such operations have the opportunity to be performed implicitly in parallel.
%Arrays and matrices within Chapel are defined over a \textit{domain}, which represents a set of indices and is treated by Chapel as a first-class construct.
%A benefit of domains is that multiple arrays can be associated with the same domain and when that domain is modified, such as resized, all of the arrays associated with that domain are logically reallocated with the new size.

% coforall
Within Chapel, parallelism can be expressed in a task or data parallel manner.
The \texttt{coforall} construct provides task-parallelism and allows users to explicitly create a specified number of concurrent tasks.
%, where each task is designated to execute one iteration of the loop.
The order in which the tasks
%iterations
are executed is non-deterministic as they are run in parallel but the statements within a given task are executed serially. % and are deterministic.
The OpenMP analogue of a \texttt{coforall} is the \texttt{omp parallel} directive.
Listing \ref{lst:coforall} provides an example of a \texttt{coforall} and Listing \ref{lst:ompParallel} shows the equivalent code in C/OpenMP, where \texttt{numTasks} and \texttt{numThreads} are equivalent.
%While not shown in the listings, Chapel provides a mechanism for task-level barriers that is equivalent to the \texttt{omp barrier} directive.

\noindent\begin{minipage}{\linewidth}
\begin{lstlisting}[style=CStyle,label={lst:coforall}, caption={Chapel coforall task-parallel construct},columns=flexible]
coforall tid in 0..numTasks-1 {
	writeln("Hello from Task ", tid);
	if tid == 0 {
		writeln("Extra hello from master: ", tid);
	}
}
\end{lstlisting}
\end{minipage}

\noindent\begin{minipage}{\linewidth}
\begin{lstlisting}[style=CStyle,label={lst:ompParallel}, caption={OpenMP equivalent of coforall construct shown in Listing \ref{lst:coforall}},columns=flexible]
#pragma omp parallel num_threads(numThreads) 
{
	int tid = omp_get_thread_num();
	printf("Hello from Task %d\n", tid);
	#pragma omp master
	{
		printf("Extra hello from master: %d\n", tid);
	}
}
\end{lstlisting}
\end{minipage}

% forall
The \texttt{forall} loop construct in Chapel allows users to express data-parallelism, where the iterations of the loop are blocked and assigned to different tasks to execute in parallel.
The number of tasks assigned to execute a given \texttt{forall} loop is a function of the iterand expression.
The OpenMP equivalent of the \texttt{forall} loop is the \texttt{omp parallel for} directive.
Listing \ref{lst:forall} shows an example of the \texttt{forall} loop and Listing \ref{lst:ompFor} provides the C/OpenMP equivalent.
In both cases, each element of \texttt{myArray} is incremented by 1 in parallel.
%Note that Chapel allows users to forgo the traditional C-style array bounds specifications in favor of array iteration. 
Furthermore, Chapel supports whole-array operations that can be performed implicitly in parallel, so the \texttt{forall} loop can be equivalently expressed with the single statement shown on line 4 in Listing \ref{lst:forall}.
%Listing \ref{lst:forall} can therefore be equivalently expressed with the single statement shown in Listing \ref{lst:broadcast}.

\noindent\begin{minipage}{\linewidth}
\begin{lstlisting}[style=CStyle,label={lst:forall}, caption={Chapel forall data-parallel loop},columns=flexible]
forall elem in myArray {
	elem += 1;
}
myArray += 1; // equivalent to above forall loop
\end{lstlisting}
\end{minipage}

\noindent\begin{minipage}{\linewidth}
\begin{lstlisting}[style=CStyle,label={lst:ompFor}, caption={OpenMP equivalent of forall loop},columns=flexible]
#pragma omp parallel for
for (int i = 0; i < arraySize; i++) {
	myArray[i] += 1;
}
\end{lstlisting}
\end{minipage}

%\noindent\begin{minipage}{\linewidth}
%\begin{lstlisting}[style=CStyle,label={lst:broadcast}, caption={Chapel array-wide broadcast operation, equivalent to Listing \ref{lst:forall}},columns=flexible]
%myArray += 1;
%\end{lstlisting}
%\end{minipage}

% sync vars
Chapel also provides a mechanism for read/write protection of variables shared among tasks.
%The \texttt{sync} qualifier can be applied to most of Chapel's primitive data types.
A variable that is declared with the \texttt{sync} qualifier is associated with a state that is either full or empty.
The state must be empty before a value can be written to the variable and the state must be full before the variable can be read. 
Attempts to access a \texttt{sync} variable in the incorrect state cause the task to block until the state changes.
The \texttt{atomic} type qualifier may also be applied to integers, reals and boolean variables.
Chapel supports the atomic operations commonly provided by other languages, such as test and set, compare, add and subtract.
While Chapel does not have a built-in lock/mutex construct, users can achieve the same effect through the use of \texttt{sync} or \texttt{atomic} variables.

% Getting C pointers
While Chapel encourages the use of high-level language features, it does provide several mechanism for lower-level C interoperability. %such as the usage of standard C types and functions.
Some of these are illustrated in Listing \ref{lst:cPtr}.
Of particular relevance to our work is the ability to retrieve a C-pointer to a Chapel array or matrix via the \texttt{c\_ptrTo} procedure, shown on line 4.
Once a C-pointer is retrieved, operations such as pointer arithmetic and pointer aliasing can be used, as shown on line 6.
%As a simple example, Listing \ref{lst:cPtr} shows the use of \texttt{c\_ptrTo} to retrieve a pointer to a Chapel matrix (line 3) and then use pointer arithmetic to access the rows of the matrix (line 5).
%The variable \texttt{myRowPtr} is a C-pointer and supports traditional array-index syntax, as shown on line 8.
Upon execution of this code, the original Chapel matrix will be modified (all elements set to 1).
%It is crucial to point out that Chapel makes no guarantees on the validity of a C-pointer to a Chapel array if that array is freed or reallocated.
%Therefore, it can be dangerous to store such pointers and reuse them throughout the program, as on Line 12.

\noindent\begin{minipage}{\linewidth}
\begin{lstlisting}[style=CStyle,label={lst:cPtr}, caption={Use of c\_ptrTo in Chapel},columns=flexible]
var rows, cols = 3;
var myDomain : domain(2) = {0..rows-1, 0..cols-1};
var myMatrix : [myDomain] int = 0;
var myPtr = c_ptrTo(myMatrix);
for row in 0..rows-1 {
	var myRowPtr = myPtr + (row * cols);
	for col in 0..cols-1 {
		myRowPtr[col] = 1;
	}
}
\end{lstlisting}
\end{minipage}

%%%%%%%%%%%%%%%%%%%%%%%%%%%%%%%%%%%%%%%%%
\section{Tensor Decomposition}
\label{sec:tensorDecomp}
%%%%%%%%%%%%%%%%%%%%%%%%%%%%%%%%%%%%%%%%%
%
% Basic intro to tensors and tensor decomposition
The need to process large amounts of sparse, high dimensional data is common in fields such as signal processing and data mining.
Analysts in these fields often need to identify previously unknown relationships among elements in the data.
\textit{Tensors}, which are matrices extended to three or more dimensions, are a natural way to model such data and \textit{tensor decomposition} is a popular technique for extracting patterns from large, multi-way data.
Tensor decomposition is often presented as the higher-order analogue of matrix singular value decomposition (SVD)~\cite{KoldaSIAM09}.
For the remainder of this paper, we refer to the dimensions of a tensor as its \textit{modes} and the number of modes in a tensor as its \textit{order}.
We refer the reader to the extensive surveys by Kolda and Bader~\cite{KoldaSIAM09} and Sidiropoulos et al.~\cite{tensorsML} for more detailed information about tensors and tensor decomposition.

% Intro CP and CP-ALS
The Canonical Decomposition/Parallel Factorization (CP) algorithm is the most commonly used approach to extend SVD to tensors. 
CP computes a set of components whose sum approximates the original tensor.
The number of components, R, is referred to as the \textit{rank} of the decomposition.
%TBR: We could get rid of this Figure if needed; just make sure we fix the text wherever we reference it
%An illustration of a rank R CP decomposition is shown in Figure \ref{fig:cpDecomp}, where $\mathcal{X}$ is a $3^{rd}$ order tensor and $\textbf{a}_{R}$, $\textbf{b}_{R}$ and $\textbf{c}_{R}$ are rank-one tensors (i.e., vectors).
% Briefly describe CP-ALS algorithm, introducing the terms that we'll need later
The most popular approach to computing these components is alternating least squares (ALS).
A basic sketch of CP-ALS is shown in Algorithm \ref{algo:CPALS} for a $3^{rd}$ order tensor $\mathcal{X}$.
While CP-ALS is general and can operate on tensors of arbitrary order, we restrict our discussion to $3^{rd}$ order tensors for simplicity.
For each mode $n$, $I_n$ denotes its length and the dense matrix $\textbf{A}^{(n)}$ represents the \textit{factor matrix} for that mode and consists of $I_n$ rows and R columns.
The columns of the factor matrices represent the rank-one components of the decomposition.
%as shown in Figure \ref{fig:cpDecomp}.
%For example, the columns of $\textbf{A}^{(1)}$ are $\textbf{a}_{1}, \textbf{a}_{2}, ..., \textbf{a}_{R}$.

It has been shown that most of the computational and storage complexity of CP-ALS stems from lines 5, 8 and 11 in Algorithm \ref{algo:CPALS}, which multiply the mode $n$ unfolded, or \emph{matricized}, tensor $\mathcal{X}^{(n)}$ with the Khatri-Rao product (written as $\odot$) of the factor matrices~\cite{rolinger2017JPDC,smith2015splatt}.
This operation is referred to as the matricized tensor times Khatri-Rao product (MTTKRP).
%Because each factor matrix has dimensions ${I_n \times R}$, the Khatri-Rao product requires O(R $\times \prod\limits_{n=1}^N I_n $) storage space, where $N$ is the order of the tensor. 
Unless the factors are very sparse, this product is totally dense due to high fill-in and can easily require many times more memory than the original tensor. 
%This considerable increase in required memory is referred to as the \emph{intermediate data explosion problem}~\cite{GigaTensor_Kang_2012}.
Due to the performance characteristics of the MTTKRP, researchers have focused on designing efficient implementations with respect to both execution time and memory~\cite{smithiaaa2015,reservoirHPEC2017}.

%
%	CP-ALS algorithm for 3rd order tensors. I want to keep the discussion to 3rd order
%	tensors as it is easier to talk about. It does make this algorithm look longer but
%	I find it more straightforward to walk through and understand.
%
\begin{algorithm}
\begin{algorithmic}[1]
\STATE \textbf{Procedure} CP-ALS($\mathcal{X}$,~R)
\STATE initialize $\textbf{A}^{(n)} \in \mathbb{R}^{I_n \times R}$ for $n$ = 1,2,3
\REPEAT 
\STATE $\textbf{V} \leftarrow \textbf{A}^{(2)T} \textbf{A}^{(2)} * \textbf{A}^{(3)T}\textbf{A}^{(3)}$ 
\STATE $\textbf{A}^{(1)} \leftarrow \textbf{$\mathcal{X}$}^{(1)}(\textbf{A}^{(3)} \odot \textbf{A}^{(2)})\textbf{V}^\dagger$
\STATE normalize columns of $\textbf{A}^{(1)}$ (storing norms as $\lambda$)
\STATE $\textbf{V} \leftarrow \textbf{A}^{(1)T} \textbf{A}^{(1)} * \textbf{A}^{(3)T}\textbf{A}^{(3)}$ 
\STATE $\textbf{A}^{(2)} \leftarrow \textbf{$\mathcal{X}$}^{(2)}(\textbf{A}^{(3)} \odot \textbf{A}^{(1)})\textbf{V}^\dagger$
\STATE normalize columns of $\textbf{A}^{(2)}$ (storing norms as $\lambda$)
\STATE $\textbf{V} \leftarrow \textbf{A}^{(1)T} \textbf{A}^{(1)} * \textbf{A}^{(2)T}\textbf{A}^{(2)}$ 
\STATE $\textbf{A}^{(3)} \leftarrow \textbf{$\mathcal{X}$}^{(3)}(\textbf{A}^{(2)} \odot \textbf{A}^{(1)})\textbf{V}^\dagger$
\STATE normalize columns of $\textbf{A}^{(3)}$ (storing norms as $\lambda$)
\UNTIL{fit ceases to improve or maximum iterations reached}
\RETURN $\lambda,~\textbf{A}^{(1)}, \textbf{A}^{(2)},\textbf{A}^{(3)}$\\
\end{algorithmic}
\caption{CP-ALS for $3^{rd}$ order tensors}
\label{algo:CPALS}
\end{algorithm}

%
%	Figure for CP decomp
%
%
%\begin{figure}
%\centering
%\includegraphics[scale=0.42, trim={0 1.75cm 0 0},clip]{Figures/CP-decomp}
%\caption{A rank $R$ CP decomposition of a $3^{rd}$ order tensor adapted from ~\cite{KoldaSIAM09}.}
%\label{fig:cpDecomp}
%\end{figure}

% What is SPLATT, where did it come from, etc.
SPLATT\footnote{SPLATT source code is available from https://github.com/ShadenSmith/splatt} is an open source software toolbox for sparse tensor factorization and related kernels~\cite{smithiaaa2015,smith2017tucker,smith2017hpc}. 
SPLATT includes routines for computing least-squares CP, as well as constrained CP and CP with missing values (i.e., tensor completion).
SPLATT is written in C and uses OpenMP+MPI for a hybrid of shared- and distributed-memory parallelism.
In this paper, we focus on SPLATT's shared-memory implementation of CP-ALS and leave the distributed implementation for future work.

% Brief description of CSF. Refer readers to Shaden's paper for a detailed breakdown of it.
SPLATT uses a novel data structure for the storage of sparse tensors, a \textit{compressed sparse fiber} (CSF), that addresses the memory/computation trade-off of computing tensor-matrix products~\cite{smithiaaa2015}.
%Instead of storing a compressed tensor for each mode of the original tensor, as is done in other CP-ALS implementations~\cite{DFacto}, CSF achieves a small memory footprint by operating on a single compressed tensor.
%To reduce the amount of additional floating point operations that are introduced by using a single compressed tensor, a novel shared-memory parallelized algorithm was developed to perform tensor-matrix multiplication on a tensor stored in CSF.
To efficiently perform MTTKRP on tensors stored in CSF, SPLATT uses a custom shared-memory parallelized algorithm.
Furthermore, CSF allows SPLATT to store and operate on tensors with an arbitrary number of modes.
We refer readers to the original publication of CSF for more details on the design of its data structure and parallelized algorithms~\cite{smithiaaa2015}.

SPLATT's approach to storing sparse tensors and its shared-memory parallel algorithm for MTTKRP has been shown to be generally the fastest with respect to execution time as well as the most efficient with regard to memory usage~\cite{rolinger2017JPDC}.

%%%%%%%%%%%%%%%%%%%%%%%%%%%%%%%%%%%%%%%%%
\section{Porting SPLATT to Chapel}
\label{sec:porting}
%%%%%%%%%%%%%%%%%%%%%%%%%%%%%%%%%%%%%%%%%
%In this section, we describe our Chapel implementation of SPLATT as well as our experience in porting the C/OpenMP code to Chapel (version 1.16).
Our goal when implementing the SPLATT code in Chapel was to simplify the code while preserving the original implementation's design and approach. Therefore, we made no significant algorithmic or data structure changes.
However, where possible, we employed Chapel's high-level language features to simplify the code.
Much of the SPLATT C code was ported to Chapel in a straightforward manner, but a few portions required a significantly different approach. In this section, we describe the challenges we encountered during porting and outline features lacking in Chapel that would have been beneficial to our work.

% Mutex pool: main take away is that we made it ourselves since it wasn't built into chapel.
%TBR: Make these into sections or not?
%##########################################
\subsection{Mutex Pool}
\label{sec:mutexPool}
%##########################################
In the C/OpenMP implementation of SPLATT, a mutex pool is used in some of the parallel MTTKRP routines.
%is created which consists of an array of \texttt{omp\_lock\_t}'s, where each \texttt{omp\_lock\_t} is padded by 64-bytes to avoid false sharing.
%When a given thread wants to update row $i$ of the output matrix, it will attempt to set lock $i$ (mod $N$), where $N$ is the size of the mutex pool.
As Chapel does not have a built-in mutex or lock, we used an array of \texttt{sync bool} variables.
%The locks are also padded to avoid false sharing.
When the pool is initialized, the \texttt{sync} variables are set to true, putting them in the "full" state, as discussed in Section \ref{sec:chplOverview}.
Acquiring a lock is equivalent to reading the \texttt{sync} variable and releasing a lock is equivalent to writing to the \texttt{sync} variable.
Similar approaches have been taken elsewhere to create arrays of OpenMP locks in Chapel~\cite{HollingsworthChapelSingle}. This technique was functionally correct, but resulted in a significant loss of performance for our application, as discussed in Section \ref{sec:mttkrp}.
Therefore, we modified our approach to use \texttt{atomic} variables instead of \texttt{sync} variables.
Listing \ref{lst:atomic} shows how acquiring and releasing a lock is implemented with \texttt{atomic} variables.

\noindent\begin{minipage}{\linewidth}
\begin{lstlisting}[style=CStyle,label={lst:atomic}, caption={Acquiring/Releasing Locks via \texttt{atomic} variables},columns=flexible]
proc set(pool : [] atomic bool, lockID : int) {
	while pool[lockID].testAndSet() {
		chpl_task_yield();
	}
}
proc unset(pool : [] atomic bool, lockID : int) {
	pool[lockID].clear();
}
\end{lstlisting}
\end{minipage}

% Omp parallel {omp for} into Chapel
%TBR: Better name for this? If there a term that describes what this is talking about?
%##########################################
\subsection{Work Sharing Constructs}
\label{sec:workSharing}
%##########################################
One pattern of parallel computation in the C SPLATT implementation involves each thread operating on its own private buffer but also iterating over a designated slice of a matrix.
This is accomplished with a \texttt{omp for} nested within a \texttt{omp parallel} section, as illustrated in Listing \ref{lst:ompIssue}.
In this code, each thread updates its own copy of \texttt{myVals} but accesses only a certain slice of rows of the matrix \texttt{vals}, as the outermost for-loop is parallelized and split up among the team of threads spawned by the \texttt{omp parallel} directive.
While a \texttt{coforall} construct can be used in place of the \texttt{omp parallel} directive, we cannot use a \texttt{forall} loop in place of the \texttt{omp for} directive.
%TBR: What this means is that each thread/task executes the forall completely independent of each other. What we want is the forall iterations that are blocked/divided up to be distributed to the already existing tasks from the coforall. 
Doing so would result in each task created by the \texttt{coforall} construct executing line 8 for all values of \texttt{i} rather than only a subset.
To the best of our knowledge, there is no direct translation of Listing \ref{lst:ompIssue} to Chapel using only its built-in constructs.
Instead, we must use a normal for-loop inside of a \texttt{coforall} construct and manually compute the loop bounds for each task.
%It is also worth noting that an implicit barrier is performed after the for-loops in Listing \ref{lst:ompIssue}, so the Chapel version needs to explicitly use a barrier to reproduce the same behavior.

\noindent\begin{minipage}{\linewidth}
\begin{lstlisting}[style=CStyle,label={lst:ompIssue}, caption={Nesting omp for within omp parallel},columns=flexible]
#pragma omp parallel 
{
	int tid = omp_get_thread_num();
	double *myVals = thdData[tid];
	#pragma omp for
	for (int i = 0; i < rows; i++) {
		for (int j = 0; j < cols; j++) {
			myVals[j] += vals[i][j] * 2;
		}
	}
	// do reduction on myVals
}
\end{lstlisting}
\end{minipage}

% Arrays of arrays where the subarrays are of a different size
%##########################################
\subsection{Array of Arrays}
\label{sec:arrayArrays}
%##########################################
Some of the data structures within the reference SPLATT implementation use an array of arrays, where the sub-arrays are of varying sizes.
This is accomplished by declaring an array of $N$ pointers, where $N$ is the number of sub-arrays, and then allocating each sub-array to be of the desired length.
While Chapel does support the concept of an array of arrays, it currently does not allow the sub-arrays to be of different sizes.
To overcome this limitation, we resorted to using an array of objects, where each object contains its own array. This approach was effective, but made the Chapel code less readable.
%TBR: Weird way to say it? I just want to say that we can make a similar data structure and it doesn't drastically change the code that we use to operate on it.
%This allowed us to create a data structure that was similar to the reference SPLATT implementation, though it does add to the overall code size and complexity.
% May not be necessary to show, but it does illustrate the added complexity.
%Listing \ref{lst:arrayArrays} shows how an array of arrays in both C and Chapel can be constructed.

%\noindent\begin{minipage}{\linewidth}
%\begin{lstlisting}[style=CStyle,label={lst:arrayArrays}, caption={Array of Arrays in C and Chapel},columns=flexible]
%/* C */
%int *arr[4];
%int rowLen = 1;
%for(int i = 0; i < 4; i++) {
%	arr[i] = malloc(sizeof(int) * rowLen);
%    rowLen++;
%}
%/* Chapel */
%class buffer {
%	var d : domain(1) = {0..1};
%    var buf : [d] int;
%}
%var arr : [0..3] buffer;
%var rowLen = 1;
%for i in 0..3 {
%	arr[i] = new buffer();
%    arr[i].d = 0..rowlen-1;
%    rowLen++;
%}
%\end{lstlisting}
%\end{minipage}

%##########################################
\subsection{Function Pointers}
\label{sec:funcPtrs}
%##########################################
In the original C/OpenMP SPLATT implementation, a pointer to a non-trivial function was passed as an argument to another function. To the best of our knowledge, this is not currently supported in Chapel. To work around this, we created a class that consists of the function we wish to pass as an argument. We then create an instance of the class and pass that object as an argument to the desired procedure.

%%%%%%%%%%%%%%%%%%%%%%%%%%%%%%%%%%%%%%%%%%%%%%%%%%%%%%%%%%%%%%%%%%
% Putting these here for now so they show up on the correct page
%%%%%%%%%%%%%%%%%%%%%%%%%%%%%%%%%%%%%%%%%%%%%%%%%%%%%%%%%%%%%%%%%%
%
% Table of the data sets and their statistics
%
\begin{table*}
\renewcommand{\arraystretch}{1.0}
\caption{Properties of Data Sets}
\label{tab:datasets}
\centering
\begin{tabular}{|c|c|c|c|c|}
\hline
\textbf{Name} & \textbf{Dimensions} & \textbf{Non-Zeros} & \textbf{Density} & \textbf{Size on Disk}\\
\hline
YELP & 41k x 11k x 75k & 8M & 1.97E-7 & 240 MB \\
RATE-BEER & 27k x 105k x 262k & 62M & 8.3E-8 & 1.85 GB \\
BEER-ADVOCATE & 31k x 61k x 182k & 63M & 1.84E-7 & 1.88 GB \\
NELL-2 & 12k x 9k x 29k & 77M & 2.4E-5 & 2.3 GB\\
NETFLIX & 480k x 18k x 2k & 100M & 5.4E-6 & 3 GB \\
\hline
\end{tabular}
\end{table*}

%
% System and Software info
%
\begin{table*}
\renewcommand{\arraystretch}{1.0}
\caption{Environment and System Properties}
\label{tab:systemVars}
\centering
\begin{tabular}{|l|l|l|}
\hline
Hardware & Software & Chapel\\
\hline
CPU: 2x E5-2697v4 Xeon Broadwell & CentOS Linux 7.4.1708 & Version: 1.16\\
Cores: 36                      & gcc 4.8.5             & Tasks: Qthreads\\
Frequency: 2.3 GHz             & OpenMP 3.1            & Memory: jemalloc\\
Last-level cache: 45 MB        & OpenBLAS 0.2.20       & flags: --fast\\
Memory: 512 GB DDR4            & SPLATT v2.0.0         & OMP\_NUM\_THREADS=1\\
\hline
\end{tabular}
\end{table*}
%
% TBR: Experiment: Make a table to show these 4 charts. Should save a lot of space
%
\begin{table*}
\renewcommand{\arraystretch}{1.0}
\caption{Runtime in seconds for CP-ALS Routines - Initial Results}
\label{tab:cpalsInitial}
\centering
\begin{tabular}{|c|c|c|c|c|c|c|c|c|}
\hline
\textbf{Data set} & \textbf{Threads/Tasks} & \textbf{Code} & \textbf{MTTKRP} & \textbf{Sort} & \textbf{Mat A\textasciicircum TA} & \textbf{Mat norm} & \textbf{CPD fit} & \textbf{Inverse} \\ 
\cline{3-9} \hline
% YELP 1 thread
\multirow{4}*{YELP} & \multirow{2}*{1}  & C & \textbf{13.31} & \textbf{0.82} & 0.34 & 0.14 & 0.04 & 0.94 \\
\cline{3-9}
&  & Chapel-initial & \textbf{225.11} & \textbf{7.21} & 0.36 & 0.14 & 0.04 & 0.98 \\
\cline{2-9}
% YELP 32 thread
& \multirow{2}*{32}  & C & \textbf{0.73} & \textbf{0.07} & 0.41 & 0.01 & 0.01 & \textbf{0.05} \\
\cline{3-9}
&  & Chapel-initial & \textbf{118.93} & \textbf{0.47} & 0.56 & 0.06 & 0.01 & \textbf{0.98} \\
\cline{3-9}
\hline
\hline
% NELL-2 1 thread
\multirow{4}*{NELL-2} & \multirow{2}*{1}  & C & \textbf{109.25}  & \textbf{7.90} & 0.13 & 0.06 & 0.01 & 0.37 \\
\cline{3-9}
&  & Chapel-initial & \textbf{1999} & \textbf{69.04} & 0.14 & 0.06 & 0.01 & 0.39 \\
\cline{2-9}
% YELP 32 thread
& \multirow{2}*{32}  & C & \textbf{5.81}  & \textbf{0.63}  & 0.24 & 0.01 & 0.01 & \textbf{0.04} \\
\cline{3-9}
&  & Chapel-initial & \textbf{88.3} & \textbf{5.01} & 0.19 & 0.02 & 0.01 & \textbf{0.39} \\
\cline{3-9}
\hline
\end{tabular}
\end{table*}

% Overall impression of Chapel; highlight things we would have wanted
%##########################################
\subsection{Commentary on Chapel Features}
\label{sec:desired Features}
%##########################################
From a programmer productivity perspective, we found porting the C/OpenMP code to Chapel to be relatively straightforward, even for inexperienced Chapel users such as ourselves.
%TBR: With the above sentence, I don't know the best way to say it. I certainly was a first-time user, but I don't want to lump you guys in there as well!
% CDK is there anything more we can add here. What about these features was helpful? Could these features have been improved to make them easier to use or even more helpful?
We particularly found features such as built-in reductions, whole array assignments and operations, array resizing and the BLAS/LAPACK interface to be of significant value for reducing code complexity.
The lack of a mutex or lock capability, having limited control over parallelization of nested loops, and the absence of native support for function pointers and jagged arrays complicated the porting effort and increased the complexity of the final source code. %Additionally, more support for functions as first class language elements would have aided our work.
%\begin{itemize}
%\item Add a mutex/lock library. As we discuss in Section \ref{sec:mutexLocks}, certain logical design choices for a mutex led to significant performance loss. If Chapel provided efficient locks, that would help others avoid these performance pitfalls.
%\item Provide more control over the behavior of \texttt{forall} loops nested within a \texttt{coforall} loop. As described above, we are unaware of any Chapel constructs that would allow a \texttt{forall} loop nested within a \texttt{coforall} loop to automatically divide work among the tasks spawned by the \texttt{coforall} loop rather than spawning new tasks.
%\item Varying sizes for sub-arrays in an array of arrays. While we were able to overcome this limitation of Chapel, it added to the overall code size and complexity, requiring a separate object to store the sub-array and its domain as well as complicating the sub-array accesses.
%\item More support for functions as first class constructs. 
%\end{itemize}

%%%%%%%%%%%%%%%%%%%%%%%%%%%%%%%%%%%%%%%%%
\section{Performance Evaluation}
\label{sec:eval}
%%%%%%%%%%%%%%%%%%%%%%%%%%%%%%%%%%%%%%%%%
% What we do in this section
Having completed our Chapel implementation of SPLATT, we turned our attention to its performance, particularly in comparison to the C implementation. Our initial Chapel code was considerably slower than the C version. 
We conducted a performance analysis to locate the most significant performance issues and identified the sorting routine, MTTKRP kernel and LAPACK-based matrix inverse routine as the code segments with the largest performance differences.
To improve the performance of these routines, we not only modified the Chapel code but also investigated the runtime environment used by Chapel and discovered significant performance issues related to the tasking layer.
%We modified the Chapel code and runtime environment to improve performance of these routines. 
Sections \ref{sec:sorting} through \ref{sec:ompConflicts} provide details on these issues and their resolutions. We present the performance of our final, optimized code in Section \ref{sec:finalResults}.
%We describe our system configuration and environment as well as the data sets we used in our evaluation.
%We present performance results that show our initial Chapel port and then discuss various optimizations and modifications that we made in order to improve its performance.

%##########################################
\subsection{System Configuration and Data Sets}
\label{sec:systemConfig}
%##########################################
% Basic info about what we did and did not include in our port.
Our Chapel code is based on SPLATT v2.0.0 (repository branch \texttt{b4bbad4-master}). Our code currently only includes SPLATT's shared-memory parallel implementation.
SPLATT's optional feature to tile the modes of a tensor was omitted from our port, as it is not commonly used, and is not evaluated in our experiments.
Also, we limited our port to operate on $3^{rd}$ order tensors.
%, even though SPLATT's framework is designed to handle tensors with an arbitrary number of modes.
This was done to simplify our port and because most tensor data sets available are $3^{rd}$ order tensors.

% Talk about data sets
We conducted performance tests on 5 different $3^{rd}$ order tensors.
% CDK -- How many other tensors did we run? If we could say "8 other tensors", it would be stronger. In fact, it might not hurt to include them in Table I, so people can see the range covered.
Across these different input tensors, we found the performance trends between the C reference implementation and our Chapel code to be similar. Therefore, for the sake of brevity, in this paper we only present the results from the YELP and NELL-2 data sets, as described in Table \ref{tab:datasets}.
% CDK Why these two and not a different two? Are these the biggest? Smallest? Densest?
The Yelp Phoenix Academic Data set, from the Yelp Dataset Challenge\footnote{https://www.yelp.com/dataset\_challenge/}, contains reviews of businesses.
The Never Ending Language Learner (NELL) data~\cite{NELL} contains subject-verb-object relationships between words.
%The NELL-2 data set was retrieved from the Formidable Repository of Open Sparse Tensors and Tools (FROSTT)~\cite{frosttdataset}.
The YELP data set is representative of small sparse tensors while the NELL-2 data set is representative of moderate to large sparse tensors.
Furthermore, the YELP data set showcases a particular feature that resulted in significant performance differences between the Chapel code and the reference C code, as discussed in Section \ref{sec:mttkrp}.

% Talk about system config
% TBR: This can be changed; perhaps just saying the basic info in the text will use less space than the table?
The details of our evaluation hardware and software are summarized in Table \ref{tab:systemVars}. 
Both the original C/OpenMP code and our Chapel code make use of the \texttt{syrk}, \texttt{potrf} and \texttt{potrs} BLAS/LAPACK routines via OpenBLAS.
Unless otherwise stated, the Chapel environment described in Table \ref{tab:systemVars} is the configuration we used.
%\texttt{QT\_AFFINITY} was set to "no" for the Chapel experiments to prevent contention between Qthread workers and the OpenMP threads used by OpenBLAS.
%This is discussed further in Section \ref{sec:inverse}.

% Talk about how we ran the experiments
For our experiments, the reference C/OpenMP code and our Chapel code performed 20 iterations of CP-ALS on each data set with a decomposition rank of 35.
The reported per-routine runtimes represent the total time for those routines at the end of the 20 iterations.
Furthermore, we performed 10 trials for both codes and report the average of those runs.
For the C/OpenMP code, we varied the number of OpenMP threads from 1 to 32 via the \texttt{OMP\_NUM\_THREADS} environment variable.
While the Chapel code also utilizes OpenBLAS and can leverage OpenMP, we opted to fix the number of OpenMP threads to 1 for our Chapel experiments.
This is further explained in Section \ref{sec:ompConflicts}.
We control the number of threads used by Qthreads within our Chapel code by setting the \texttt{CHPL\_RT\_NUM\_THREADS\_PER\_LOCALE} environment variable.
This restricts both the C/OpenMP and Chapel code to use the same number of threads during each run.
Otherwise, Qthreads defaults to a thread count that is equal to the number of cores on the system.
To control the number of tasks used for the \texttt{coforall} constructs, we varied a user-level configuration variable that is used in the loop iterator.

%
%	This section presents the initial results (i.e. C vs unoptimized Chapel).
%	The purpose is to set up our motivation for the optimizations
%
%##########################################
\subsection{Initial Results}
\label{sec:initialResults}
%##########################################
Table \ref{tab:cpalsInitial} shows the execution times of the main routines for the C and Chapel code on the YELP and NELL-2 data sets when using 1 and 32 threads/tasks.
Inverse refers to computing the Moore-Penrose inverse (denoted as $\textbf{V}^\dagger$ in Algorithm \ref{algo:CPALS}), Mat A\textasciicircum TA refers to performing lines 4, 7, and 10 in Algorithm \ref{algo:CPALS}, Mat norm refers to normalizing the columns of the factor matrices (lines 6, 9, and 12), CPD fit refers to computing the fit of the decomposition (line 13) and Sort refers to the pre-processing step that SPLATT performs to sort the non-zeros in the tensor prior to performing CP-ALS.

%The two routines that generally contribute the most to the overall runtime are sorting and MTTKRP, which are bold-faced in Table \ref{tab:cpalsInitial}. 
%Unfortunately, these are the two routines in which our Chapel code performs the worst relative to the C code.
The routines that exhibit the most severe performance issues for our Chapel code are sorting, MTTKRP, and inverse (shown in bold-face in Table \ref{tab:cpalsInitial}).
MTTKRP and sorting in Chapel are about 18x and 8.7x slower than the C code, respectively.
It is also clear that there is a scalability issue on the YELP data set for the Chapel MTTKRP, as it only achieves a 1.9x speed up from 1 to 32 tasks.
%There is also a clear performance difference between the C and Chapel code for the inverse routine when using 32 threads/tasks.
Furthermore, as mentioned in Section \ref{sec:systemConfig}, we executed our Chapel experiments with \texttt{OMP\_NUM\_THREADS} set to 1, preventing any parallelization of the inverse procedure that is implemented via LAPACK routines from OpenBLAS.
Our reasoning for doing this is explained in Section \ref{sec:ompConflicts}.
In the following sections, we investigate these routines and present modifications to the Chapel code to improve performance.

%##########################################
\subsection{Sorting Optimizations}
\label{sec:sorting}
%##########################################
As a pre-processing step, SPLATT employs a parallel counting sort, which consists of a recursive quicksort routine that is modified to specifically sort non-zeros from $3^{rd}$ order tensors.
We profiled our Chapel code to identify what portions of this sorting routine were consuming the most time.
We identified one line of code in the quicksort routine that declared a local auxiliary array of two integers.
%This array is local to each recursive call, so it is created each time the function is entered.
In the case of the NELL-2 data set when using one thread or task, a total of 46 million recursive calls are made to the quicksort routine, and thus 46 million array allocations.
While creating an array of that size by itself exhibits very little overhead, it is still a more expensive task than the equivalent C code since Chapel arrays are high-level constructs.
We found that this reoccurring array creation can account for as much as 10\% of the sorting runtime. To mitigate this, we simply declared two separate integer variables rather than an array of two integers, requiring only minimal changes to the code.

The other portion of the code, which contributed the most to the sorting runtime, was a small loop executed before the quicksort routine was performed.
For two array of arrays, $A$ and $B$, that are of length $N$ where $N$ is the order of the tensor, the loop reassigns the sub-arrays of $A$ to the corresponding sub-arrays of $B$.
The length of each sub-array is equal to the number of non-zeros in the tensor, which can be significant in the case of large tensors.
In C, the sub-arrays are pointers, making this a simple assignment operation.
In our initial Chapel code, we implemented each array of arrays as a 2D matrix and performed the sub-array assignment using slicing.
%, since each sub-array was of equal size, and performed the sub-array assignment using slicing. 
As slicing can create excessive overhead (see Section \ref{sec:mttkrp}), we opted to use an array of arrays rather than a 2D matrix.
However, in reassigning each sub-array, Chapel performs a copy of one sub-array to another while the C code is just reassigning pointers.
To match this behavior, we used the \texttt{c\_ptrTo} procedure in Chapel to obtain pointers to the main arrays and then used the equivalent C syntax to reassign the sub-arrays.
This change improved the entire sorting routine by roughly 4x on both the YELP and NELL-2 data set.

\begin{figure}
\centering
\includegraphics[scale=0.36]{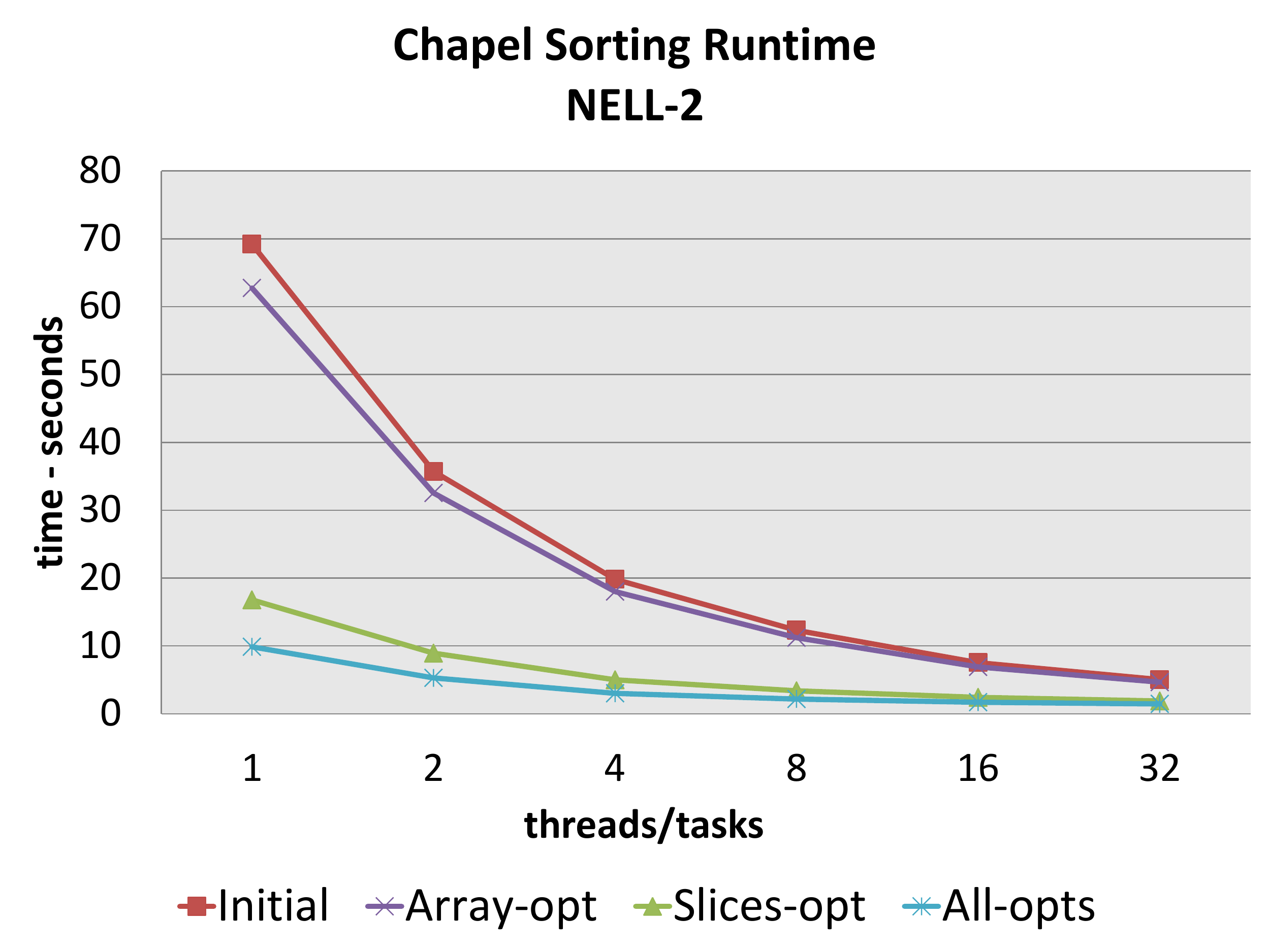}
\caption{Chapel sorting runtime on the NELL-2 data set when using 1 to 32 threads/tasks.
Initial refers to our unoptimized Chapel code, Array-opt refers to the optimization that eliminates the creation of a small integer array in a frequently called function. Slices-opt refers to the optimization that eliminates the use of slicing in array reassignment operations and All-opts refers to the Chapel code when using both Array-opt and Slices-opt. All runtimes are shown in seconds.}
\label{fig:sortNell}
\end{figure}

Combining both the array creation and slicing modifications mentioned above, we improved the sorting routine by as much as 8x from our original version.
Figure \ref{fig:sortNell} shows how the different modifications affected the overall sorting runtime for the NELL-2 data set as the number of threads/tasks are varied from 1 to 32.
We observed similar trends for the YELP data set.

%
%	Sorting Runtimes on Nell-2
%
%

%##########################################
\subsection{MTTKRP Optimizations}
\label{sec:mttkrp}
%##########################################
As the MTTKRP is the critical routine of CP-ALS, we focused heavily on improving its performance within our Chapel code.
Through our performance profiling, we identified two areas that contribute the most to the performance of  our implementation: matrix row accessing and mutexes/locks.

%@@@@@@@@@@@@@@@@@@@@@@@@@@@@@@@@@@@@@@@@@@@@@@@
\subsubsection{Matrix Row Accesses}
\label{sec:matRows}
%@@@@@@@@@@@@@@@@@@@@@@@@@@@@@@@@@@@@@@@@@@@@@@@
%As eluded to in Section \ref{sec:sorting}, array slicing can present significant overhead.
A frequent pattern in the MTTKRP code is to obtain a pointer to a particular row in a factor matrix and then perform a multiply/accumulate across the elements in the row, similar to the code shown in Listing \ref{lst:cPtr}.
In the C implementation of SPLATT, the factor matrices are stored as 1D arrays in row-major order, so accessing any given row can be done simply through pointer arithmetic.
%Obtaining a reference to a particular row $i$ is done simply by offsetting the 1D array by $i*cols$ where $cols$ is the number of columns in the factor matrix (i.e., the rank of the decomposition).
Our Chapel code represents the factor matrices as 2D matrices, with the goal of leveraging Chapel's built-in features such as slicing.

In our initial code, we obtained a row reference by slicing the factor matrix.
%and storing the result as a \texttt{ref} variable, which stores a reference rather than making a copy.
An issue with using slicing in this manner in our code is that the amount of computation that is performed on each slice is negligible when compared to the overhead of obtaining the slice. Each slice only consists of $R$ elements, where $R$ is the rank of the decomposition and typically small (35 in our case), and the computation is fairly light.
As described in a Chapel GitHub issue\footnote{https://github.com/chapel-lang/chapel/issues/8203}, array slicing can be expensive due to computing and creating the domain of the resulting array view and creating and setting up the array descriptor for the view.

Our first approach was to eliminate slicing by using direct 2D indexing for matrices, even though it deviated from the reference implementation of SPLATT.
We found that this modification led to as much as a 12x and 17x speed-up for the YELP and NELL-2 MTTKRP runtimes, respectively.
Next, we adopted a more direct translation of the C code into our Chapel code by obtaining a pointer to the factor matrices and then using pointer arithmetic to access the particular rows.
This change produced about a 1.26x speed-up over the 2D indexing approach.
Figures \ref{fig:matrixOptsYelp} and \ref{fig:matrixOptsNell} present these optimizations for the YELP and NELL-2 data sets, respectively, when varying the number of threads/tasks from 1 to 32.

%@@@@@@@@@@@@@@@@@@@@@@@@@@@@@@@@@@@@@@@@@@@@@@@
\subsubsection{Mutexes/Locks}
\label{sec:mutexLocks}
\begin{figure}
\centering
\includegraphics[scale=0.36]{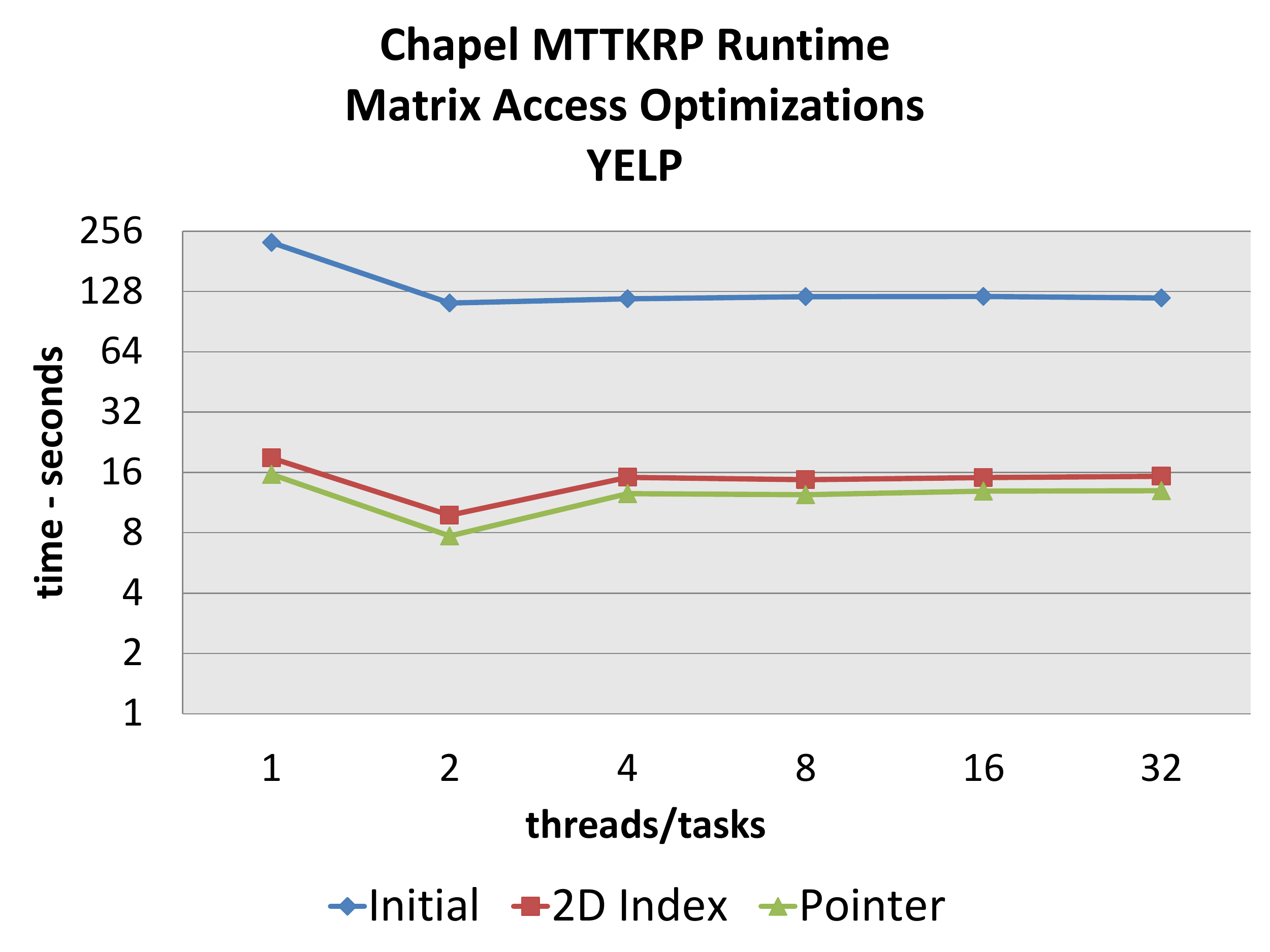}
\caption{Chapel MTTKRP runtime on the YELP data set when using 1 to 32 threads/tasks. Initial refers to our unoptimized Chapel code using slicing, 2D Index refers to replacing slicing with more direct 2D indexing and Pointer refers to retrieving C-pointers and using pointer arithmetic. All runtimes are shown in seconds and the vertical axis is logarithmic.}
\label{fig:matrixOptsYelp}
\end{figure}

\begin{figure}
\centering
\includegraphics[scale=0.36]{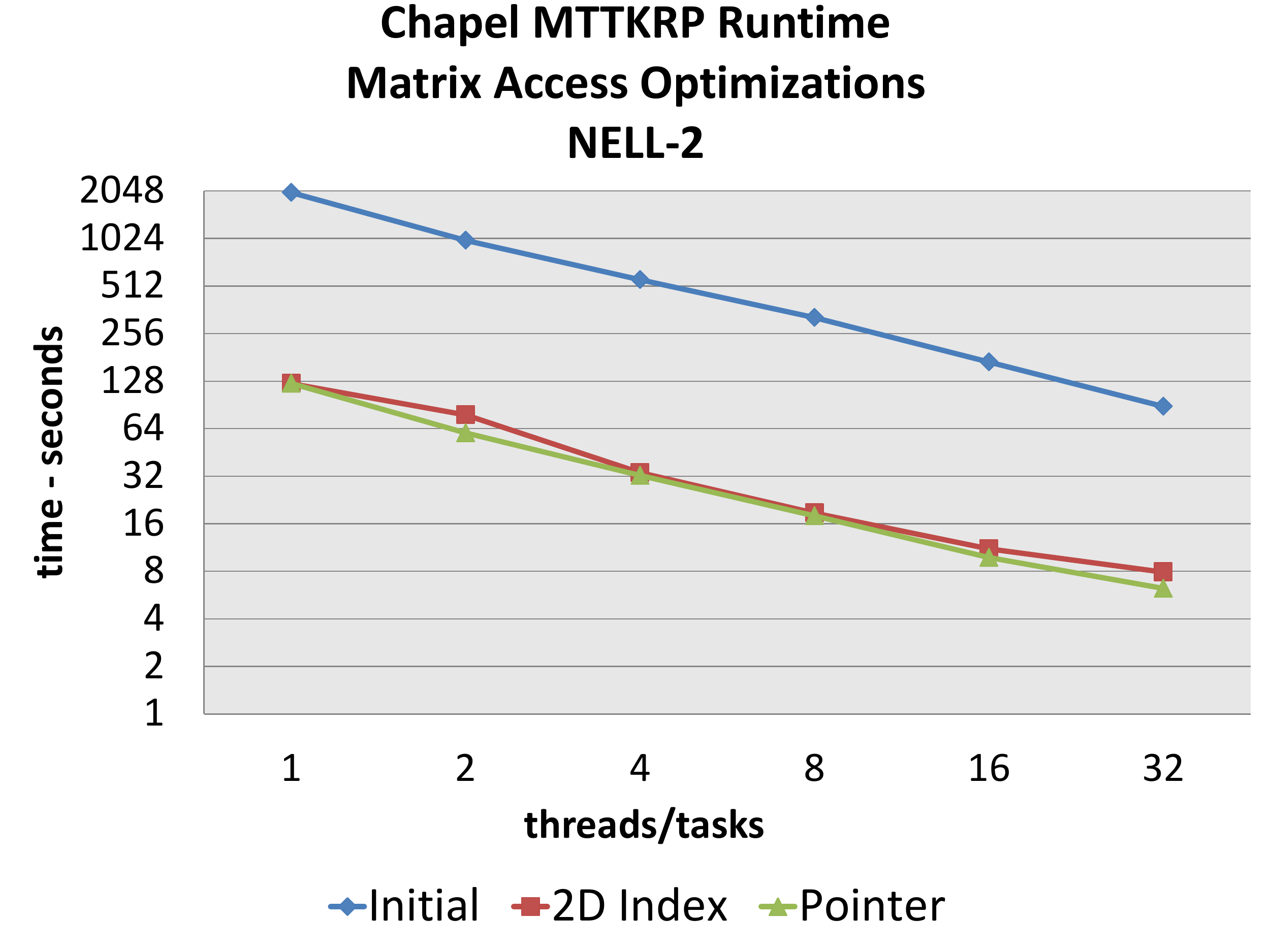}
\caption{Chapel MTTKRP runtime on the NELL-2 data set when using 1 to 32 threads/tasks. Initial refers to our unoptimized Chapel code using slicing within the MTTKRP, 2D Index refers to replacing slicing with more direct 2D indexing and Pointer refers to retrieving C-pointers and using pointer arithmetic. All runtimes are shown in seconds and the vertical axis is logarithmic.}
\label{fig:matrixOptsNell}
\end{figure}

%@@@@@@@@@@@@@@@@@@@@@@@@@@@@@@@@@@@@@@@@@@@@@@@
It is obvious from Figure \ref{fig:matrixOptsYelp} that the MTTKRP runtime for the YELP data set is exhibiting poor scalability. This behavior is not seen with the reference C/OpenMP implementation.
However, the results for the NELL-2 data set in Figure \ref{fig:matrixOptsNell} do not show the same poor scalability, rather showing near linear speed-up.
We noticed that the main difference between the YELP and NELL-2 data sets with respect to the MTTKRP is that for all thread/task counts beyond two for the YELP data set, the SPLATT algorithm will require the use of locks during the MTTKRP, as briefly mentioned in Section \ref{sec:porting}, while the NELL-2 data set will perform ``no-lock" versions of the MTTKRP for all thread/task counts.
The decision of whether or not to use locks is highly dependent on the tensor and the number of threads being used.

%
%	Matrix MTTKRP Optimizations on YELP
%
%
%
%	Matrix MTTKRP Optimizations on NELL-2
%
%
%
%	Sync vs Atomic on YELP
%
%
\begin{figure}
\centering
\includegraphics[scale=0.36]{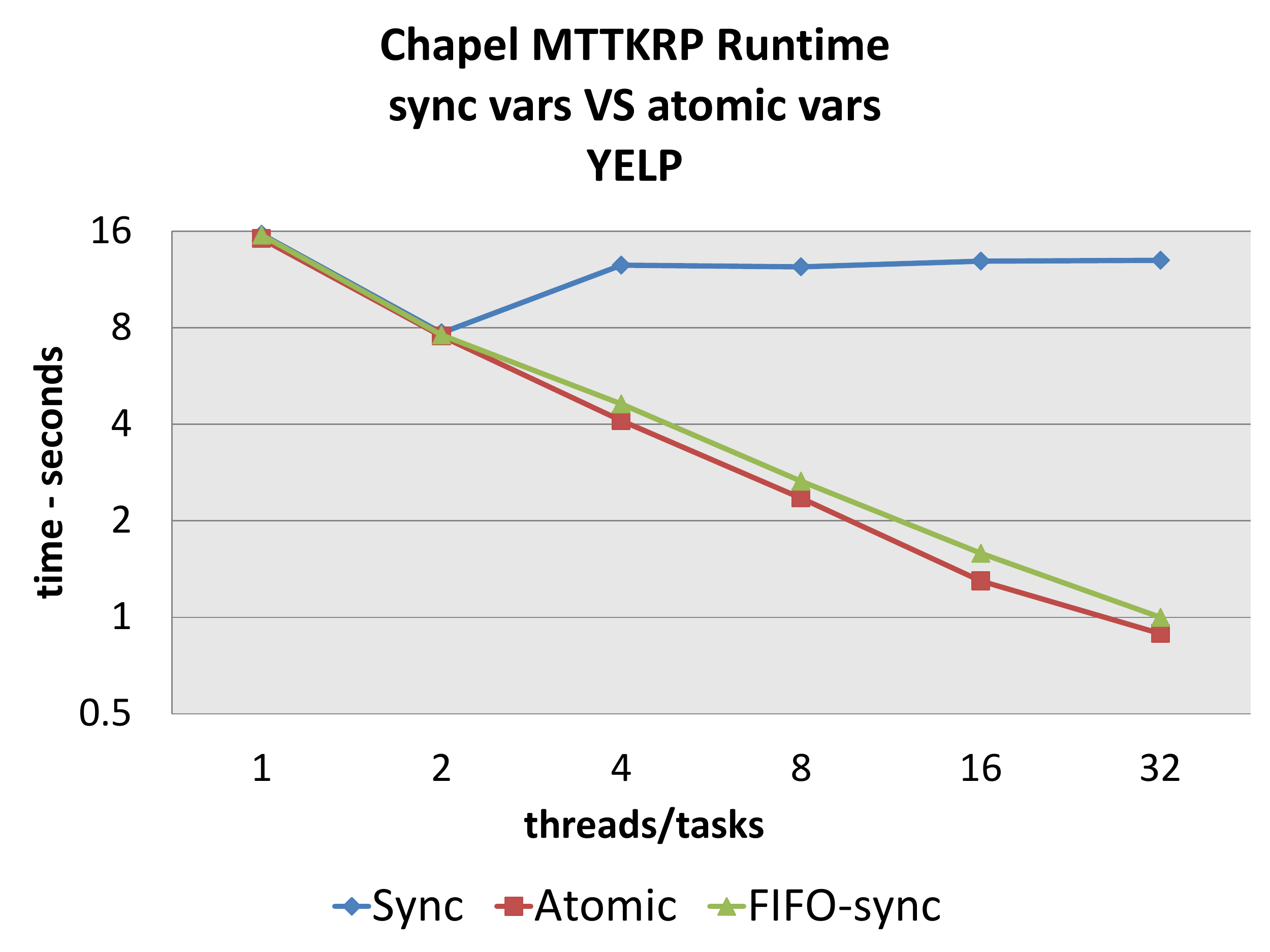}
\caption{Chapel MTTKRP runtime on the YELP data set when using 1 to 32 threads/tasks. Sync refers to using \texttt{sync} variables for our mutex pool implementation, Atomic refers to using \texttt{atomic} variables for our mutex pool implementation and FIFO-sync refers to using \texttt{sync} variables but with the \texttt{fifo} tasking layer. All runtimes are shown in seconds and the vertical axis is logarithmic. All implementations shown use the \texttt{Pointer} optimization for accessing the factor matrices within the MTTKRP.}
\label{fig:syncAtomicYelp}
\end{figure}

%SPLATT will privatize some modes of the tensor to avoid lock contention during the MTTKRP.
%This decision is based on the length of the mode, the number of threads being used, and the number of non-zeros in the tensor.
%It can be formally expressed as $length * numThreads \leq thresh * nnz$, where $thresh$ is a constant with a default value of 0.02.
%If that statement is true, then the mode will be privatized and locks will not be used when performing the MTTKRP for that mode.

In our initial code, locks were implemented with \texttt{sync} variables.
We discussed our implementation with the Chapel team and they suggested using \texttt{atomic} variables instead, as \texttt{sync} variables can be very expensive per operation, especially within the Qthreads tasking layer.
Within Qthreads, \texttt{sync} variables are better suited for heavily contended locks held for long periods of time. This is because if the variable is locked when requested, the requesting task will be put to sleep rather than execute a spin-wait.
However, the critical sections in the MTTKRP that are protected by locks consist of very short, non-intensive computation.
%Specifically, the critical sections amount to updating a row of the MTTKRP output matrix, which only consists of $R$ elements, where $R$ is the rank of the decomposition (35 in our experiments).
Using this information, we modified our mutex implementation to use \texttt{atomic} variables, as described in Section \ref{sec:porting}, which employ a spin-wait behavior that is better suited for our application.
By doing so, we significantly improved both the scalability and overall MTTKRP runtime for the YELP data set, as shown in Figure \ref{fig:syncAtomicYelp}.
%Note that the implementations shown in this figure use the \texttt{Pointer} optimization for matrix accesses, as discussed above.
As the NELL-2 data set does not require the use of locks, the performance of \texttt{sync} vs \texttt{atomic} variables is irrelevant.
%TBR: Probably unnecessary sentence below, but it is true
The particular case of the YELP data set speaks to the high irregularity in the sparse tensor decomposition application and the issues that can arise due to over-optimization for a particular data set.

As part of our investigation into the MTTKRP runtime and the mutex implementation, we tried using an alternative tasking layer to Qthreads, namely fifo (i.e., POSIX threads).
Figure \ref{fig:syncAtomicYelp} includes the results for using fifo and \texttt{sync} variables, which is competitive with the Qthreads and \texttt{atomic} implementation.
This is because under the fifo tasking layer, \texttt{sync} variables are implemented with a spin-wait like behavior rather than putting the task to sleep.
%While spin-waiting would result in poor performance for long-held locks, they perform well in our particular case.
This observation underscores the importance of the tasking layer used and its effects on Chapel's features.

%
%	CP-ALS Routines for YELP with 1 thread/task (FINAL)
%
%
\begin{figure}
\centering
\includegraphics[scale=0.36]{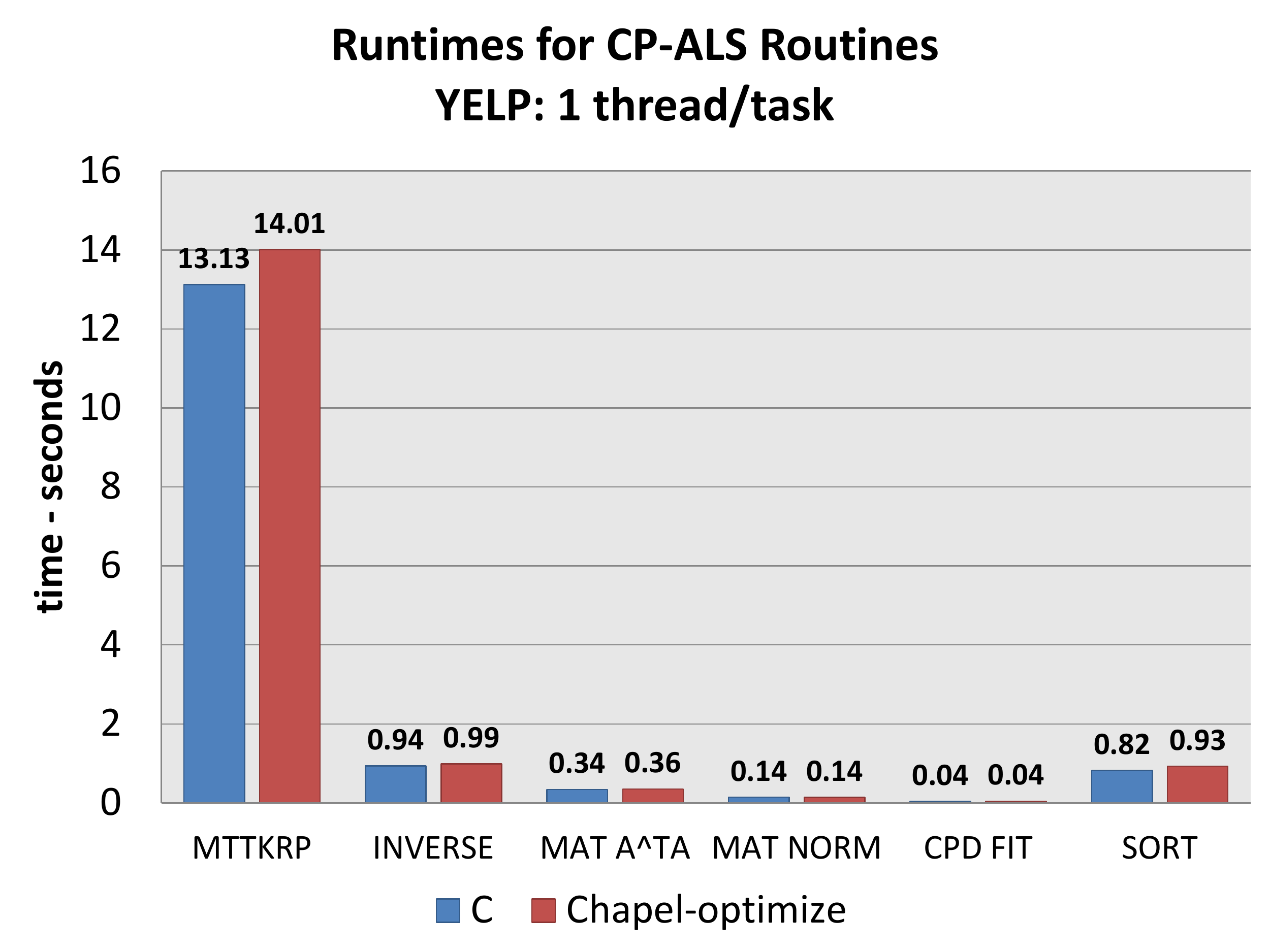}
\caption{Runtimes for CP-ALS routines on the YELP data set when using 1 thread/task. Times are shown in seconds. Lower bars represent better performance. C refers to the original C/OpenMP code and Chapel-optimize refers to our fully optimized Chapel code (i.e., all sorting optimizations, C-pointers within MTTKRP and \texttt{atomic} variables for mutexes).}
\label{fig:routinesFinalYelp1}
\end{figure}

%
%	CP-ALS Routines for NELL-2 with 1 thread/task (FINAL)
%
%
\begin{figure}
\centering
\includegraphics[scale=0.36]{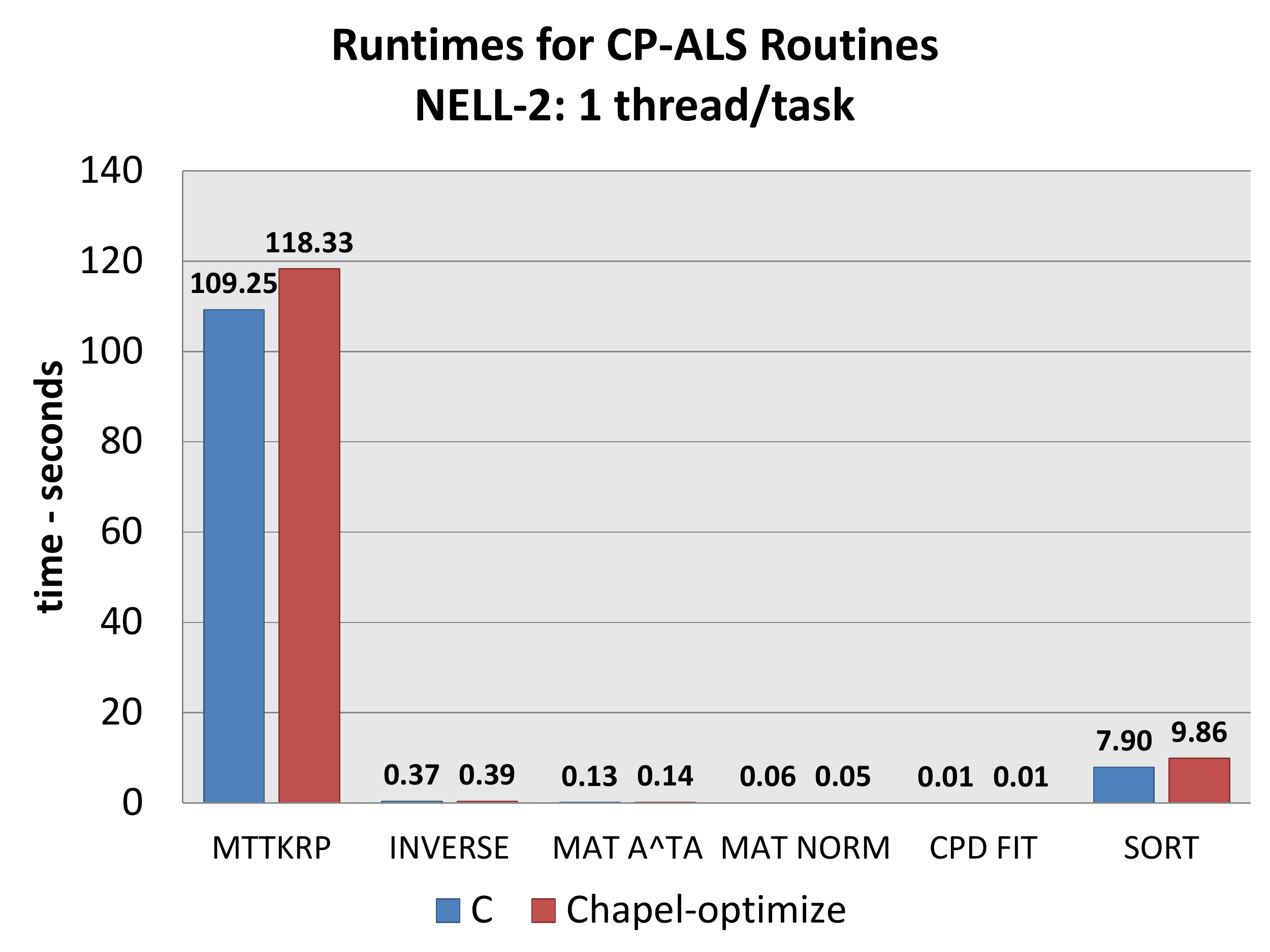}
\caption{Runtimes for CP-ALS routines on the NELL-2 data set when using 1 thread/task. Times are shown in seconds. Lower bars represent better performance. C refers to the original C/OpenMP code and Chapel-optimize refers to our fully optimized Chapel code (i.e., all sorting optimizations, C-pointers within MTTKRP and \texttt{atomic} variables for mutexes).}
\label{fig:routinesFinalNell1}
\end{figure}

%
%	CP-ALS Routines for YELP with 32 thread/task (FINAL)
%
%
\begin{figure}
\centering
\includegraphics[scale=0.36]{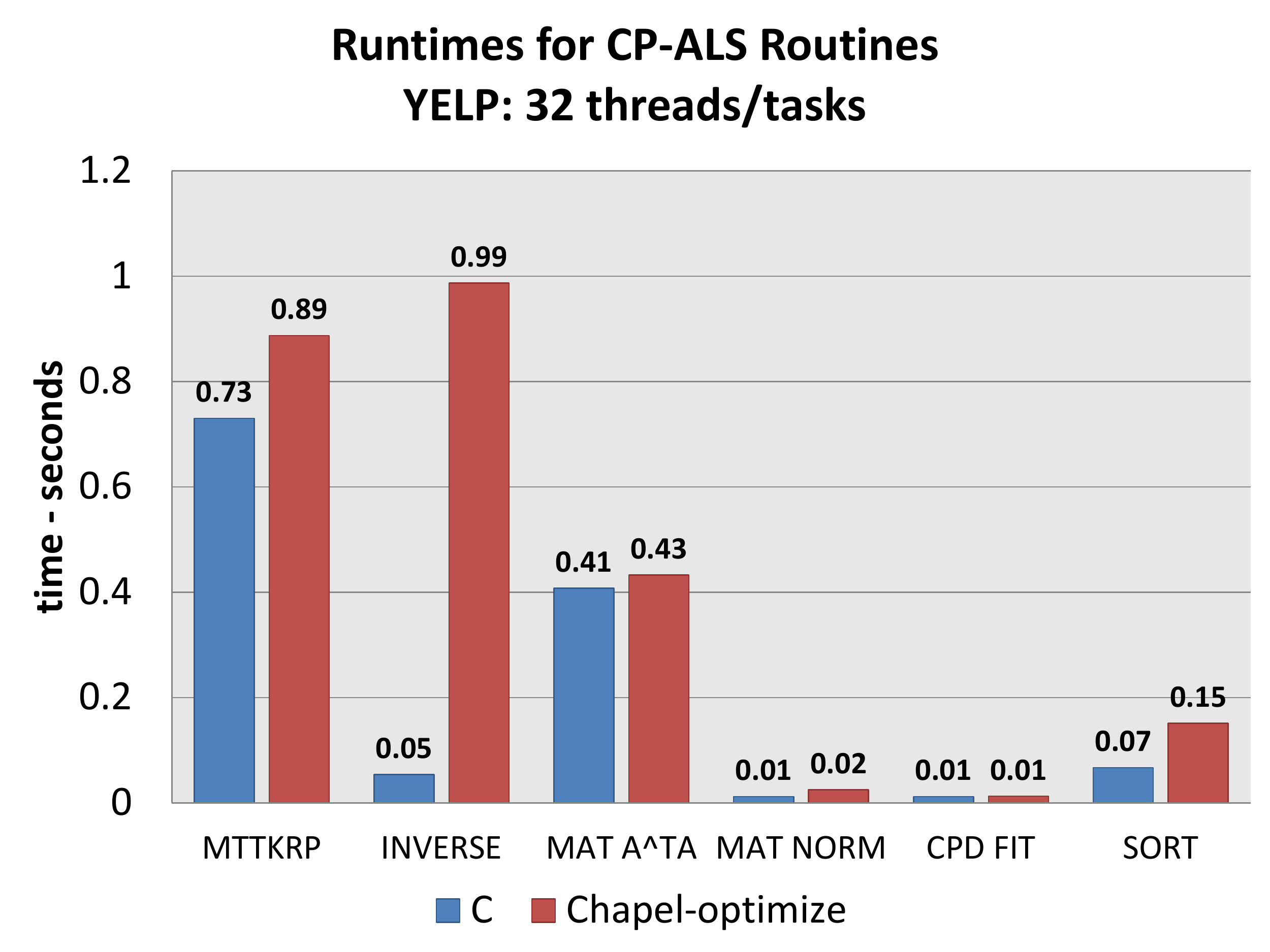}
\caption{Runtimes for CP-ALS routines on the YELP data set when using 32 threads/tasks. Times are shown in seconds. Lower bars represent better performance. C refers to the original C/OpenMP code and Chapel-optimize refers to our fully optimized Chapel code (i.e., all sorting optimizations, C-pointers within MTTKRP and \texttt{atomic} variables for mutexes).}
\label{fig:routinesFinalYelp32}
\end{figure}

%
%	CP-ALS Routines for NELL-2 with 32 thread/task (FINAL)
%
%
\begin{figure}
\centering
\includegraphics[scale=0.36]{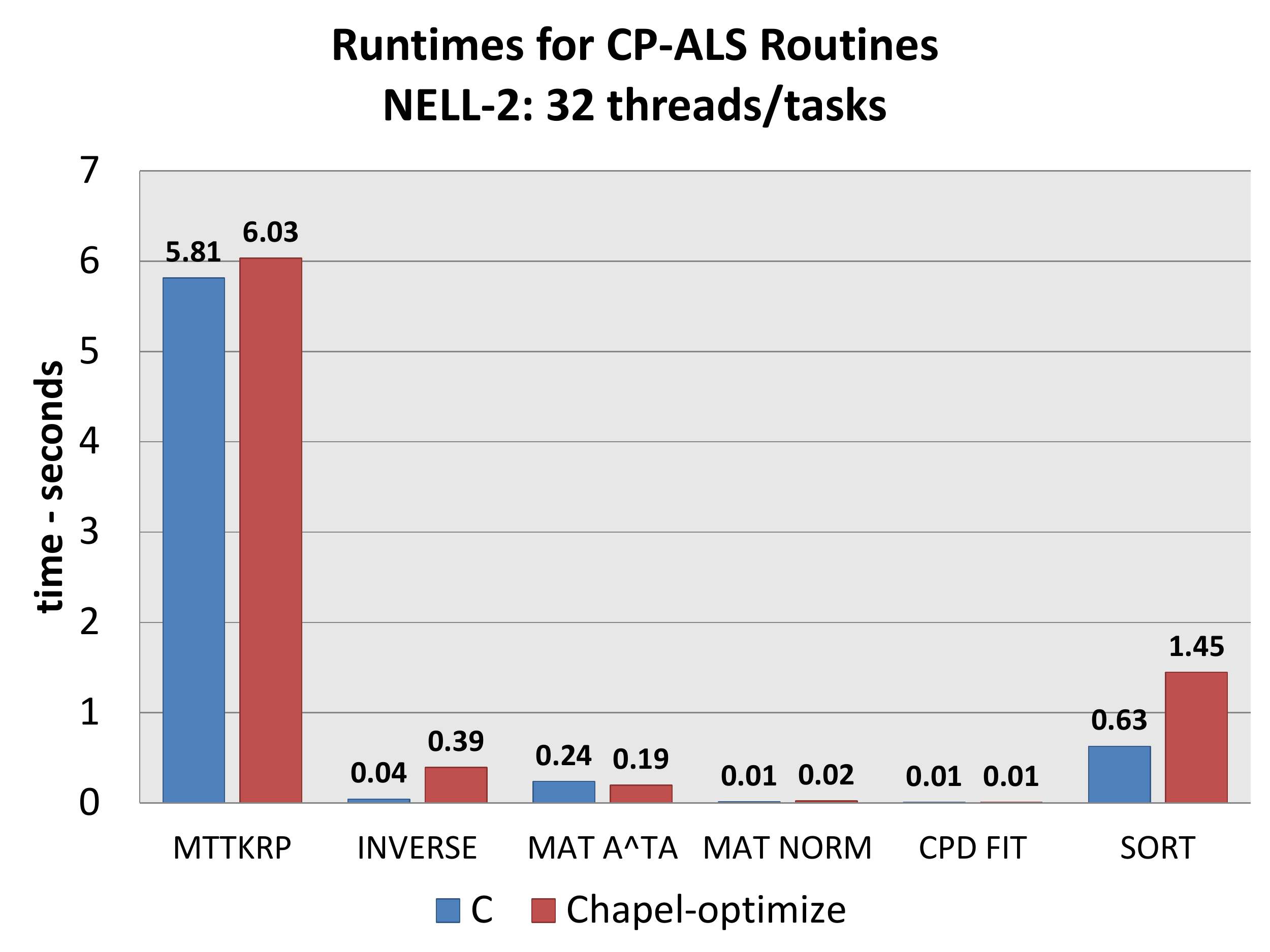}
\caption{Runtimes for CP-ALS routines on the NELL-2 data set when using 32 threads/tasks. Times are shown in seconds. Lower bars represent better performance. C refers to the original C/OpenMP code and Chapel-optimize refers to our fully optimized Chapel code (i.e., all sorting optimizations, C-pointers within MTTKRP and \texttt{atomic} variables for mutexes).}
\label{fig:routinesFinalNell32}
\end{figure}
% TBR: Better name for section?
%##########################################
\subsection{Conflicts Between Qthreads and OpenMP}
\label{sec:ompConflicts}

In both the C and Chapel codes, the matrix inverse procedure is computed using two LAPACK routines provided by OpenBLAS, which are parallelized via OpenMP. When running our Chapel code with multiple OpenMP threads, we encountered performance issues for not only the inverse procedure but other routines using QThreads. We have isolated these issues to conflicts between the Qthreads tasking layer and the OpenMP threads used by OpenBLAS.

As we increased the number of OpenMP threads from 1 to 32, we observed that the inverse routine's runtime on the YELP data set became 15x slower at 32 threads than the serial case. By default, Qthreads pins workers to cores for optimal affinity. This can negatively impact OpenMP threads by using resources on cores for overhead tasks such as checking for more work. We tried setting \texttt{QT\_AFFINITY=no}, which allows spin-waiting threads to migrate to different cores that are out of the way of the OpenMP threads. We found that this drastically improved the performance of the matrix inverse routine, achieving a 2x speed-up rather than the initial 15x slow down. However, this runtime is still roughly 10x slower than the C code for the YELP data set.

%We believe one factor hurting OpenMP performance is the amount of spin-waiting that Qthreads workers perform while waiting for a new task to be queued before being suspended.
To further reduce the conflicts between OpenMP and QThreads, we reduced the spin-wait interval. By default, Qthreads workers will spin-wait for 300,000 iterations before suspending. While this spin-waiting is occurring, it can degrade the performance of the OpenMP threads sharing those cores. In our case, the contention is large because the inverse routine has a relatively short runtime and follows a section of QThreads activity. With a long spin-wait period, a significant spin-wait overlap occurs. Following the suggestion in a Chapel GitHub issue\footnote{https://github.com/chapel-lang/chapel/issues/8337}, we shortened the amount of spin-waiting by setting \texttt{QT\_SPINCOUNT=300}. Doing so further improved the performance of the inverse procedure on 32 threads by 2.3x. However, even this result is still 4x slower than the C code.

Just as QThreads can degrade OpenMP performance, the opposite is also true. Using \texttt{QT\_AFFINITY=no} resulted in unpredictable performance when using 32 threads for the matrix normalization routine, which is written entirely in Chapel and follows directly after the OpenMP-based inverse procedure. We observed a 7x -- 13x slow down in the matrix normalization runtime at 32 threads when compared to our default configuration. We theorize that, without QThreads core pinning, spin-waiting QThreads and OpenMP threads migrate away from one another. Once the program exits the OpenMP region, these QThreads will re-migrate back out to idle cores if necessary. The contention is minor when the system has enough cores to accommodate all QThreads and OpenMP threads. When using 32 threads, a majority of the cores in our 36 core system are being used by the OpenMP threads during the inverse procedure and by the Qthread workers during the matrix normalization routine, leading to significant migration with its accompanying cache rewarming and context switch overhead. 

There does not appear to be a clear solution to overcoming these issues. If there were BLAS routines implemented entirely in Chapel, that could avoid the conflicts between Qthreads and OpenMP. Because the inverse routine generally contributes a small amount to the overall CP-ALS runtime and is the only procedure in our code that can benefit from using OpenMP threads, we have opted to set the number of OpenMP threads to 1 for our Chapel experiments.

%##########################################
\subsection{Final Results}
\label{sec:finalResults}
%##########################################
%
%	MTTKRP Runtimes on YELP (final)
%
%
\begin{figure}
\centering
\includegraphics[scale=0.36]{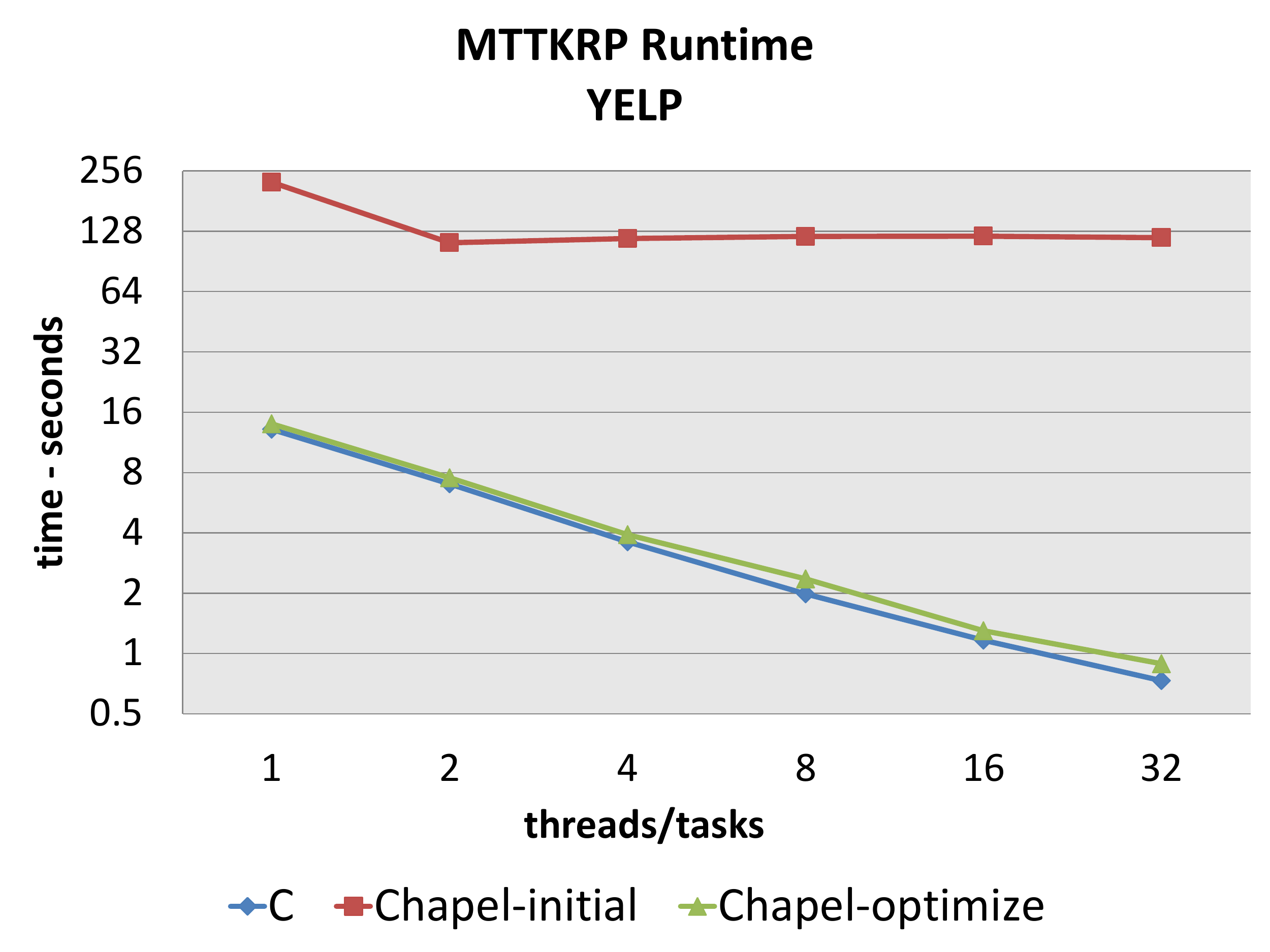}
\caption{MTTKRP runtime on the YELP data set when using 1 to 32 threads/tasks. C refers to the original C/OpenMP code, Chapel-initial refers to our original unoptimized version of the Chapel code and Chapel-optimize refers to our fully optimized Chapel code (i.e., all sorting optimizations, C-pointers within MTTKRP and \texttt{atomic} variables for mutexes). The vertical axis is logarithmic and shown in seconds.}
\label{fig:mttkrpFinalYelp}
\end{figure}

%
%	MTTKRP Runtimes on NELL-2 (final)
%
%
\begin{figure}
\centering
\includegraphics[scale=0.36]{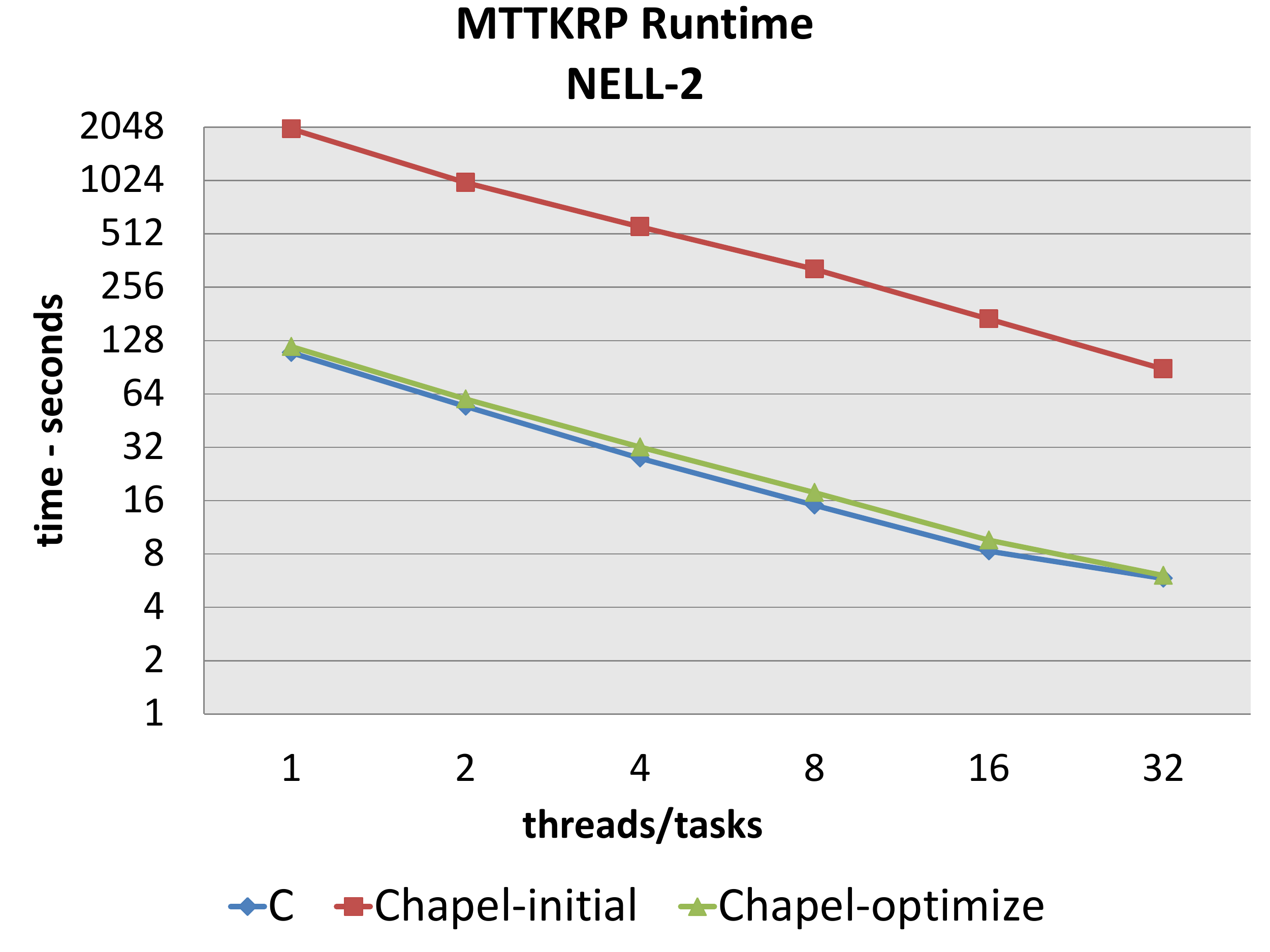}
\caption{MTTKRP runtime on the NELL-2 data set when using 1 to 32 threads/tasks. C refers to the original C/OpenMP code, Chapel-initial refers to our original unoptimized version of the Chapel code and Chapel-optimize refers to our fully optimized Chapel code (i.e., all sorting optimizations, C-pointers within MTTKRP and \texttt{atomic} variables for mutexes). The vertical axis is logarithmic and shown in seconds.}
\label{fig:mttkrpFinalNell}
\end{figure}

Figures \ref{fig:routinesFinalYelp1} and \ref{fig:routinesFinalNell1} present the per-routine runtimes of the reference C/OpenMP implementation and our fully optimized Chapel implementation when using one thread/task on the YELP and NELL-2 data sets, respectively.
We also show the 32 threads/tasks results in Figures \ref{fig:routinesFinalYelp32} and \ref{fig:routinesFinalNell32}.
Furthermore, the MTTKRP runtime on the YELP and NELL-2 data sets as the number of threads/tasks is varied from 1 to 32 are presented in Figures \ref{fig:mttkrpFinalYelp} and \ref{fig:mttkrpFinalNell}, respectively.

We observe drastic improvements with respect to our original Chapel code (Table \ref{tab:cpalsInitial}) for the MTTKRP and sorting routines.
While our final Chapel code's performance for sorting is still slightly worst than the C/OpenMP code and could be investigated further for more improvement, it only represents a small portion of the overall CP-ALS runtime.
It is evident that our Chapel code achieves competitive parallel performance and scalability for the MTTKRP when compared to the reference C/OpenMP implementation.
Overall, the Chapel code achieves 83\%-93\% of the MTTKRP performance of the C/OpenMP code on the YELP data set and 84\%-96\% of the MTTKRP performance on the NELL-2 data set. 
It also achieves near linear scalability up to 32 cores on both data sets.
However, it is clear that the inverse routine suffers in performance (most notably in Figure \ref{fig:routinesFinalYelp32}) due to the parallelization issues mentioned in Section \ref{sec:ompConflicts}.

%%%%%%%%%%%%%%%%%%%%%%%%%%%%%%%%%%%%%%%%%
\section{Related Work}
\label{sec:related}
%%%%%%%%%%%%%%%%%%%%%%%%%%%%%%%%%%%%%%%%%
% Tensor related work; key take-away: many different tensor codes out there, but none implemented in a language like Chapel.
Designing and implementing sparse parallel tensor decomposition algorithms
% for high performance systems
has been a rich research area in recent years.
There are several shared-memory based tensor decomposition implementations using novel approaches to performing MTTKRP in a scalable and efficient manner~\cite{smithiaaa2015,smithIPDPS2017,reservoirHPEC2017}.
However, to the best of our knowledge, the work presented in this paper is the first to implement parallel sparse tensor decomposition in a high productivity programming language such as Chapel.
%Such implementations have also been adopted to distributed-memory systems, where much of the research centers on optimal distribution schemes for the original tensor as well as the factor matrices~\cite{smith2016medium,DFacto,Kaya:2015:SST:2807591.2807624}.
%Similarly, graphics processing units (GPUs) have been leveraged for tensor decomposition for both single-GPU~\cite{GPUCluster17} and distributed multi-GPU architectures~\cite{rolingerHPEC2017}.

%Evaluating and understanding the performance of these different implementations with respect to tensor properties has also been an active area of research~\cite{rolinger2017JPDC}.

% Chapel related work 
There has been a significant effort to evaluate and analyze the performance of Chapel programs for both single- and multi-node environments.
Johnson and Hollingsworth ported and optimized several C/OpenMP based benchmarks to single-node Chapel including LULESH, MiniMD, and CLOMP~\cite{HollingsworthChapelSingle}.
Haque and Richards implemented an optimized multi-node version of CoMD in Chapel as well as identified key limitations of Chapel in regards to scope-based code locality~\cite{chapelCoMD}. Our work, while similar, differs from these efforts in that SPLATT is a sparse application from a different problem domain. It also is a full application with several components, ranging from file I/O and sorting to custom sparse data structures and parallel algorithms, rather than a benchmark or proxy application.

There has also been work on developing techniques to more effectively measure the performance of Chapel programs, where a data-centric view of performance data is studied as opposed to more traditional code-centric views~\cite{dataCentric}.
In our work, we employed code-centric profiling to identify performance bottlenecks via gprof and source-code level timers.
In the future, we would be interested in applying such data-centric techniques to improve our code.
%CDK We need to tie this work to what we did
%
%TBR: Assuming you're referring to the code-centric related work, then we can say that our profiling approach was data-centric (i.e. we looked at which functions or statements took the most time), instead of looking at which data-objects consumed the most time (as in, the operations performed on myArr[] contributed to 80% of the time of my program).

%%%%%%%%%%%%%%%%%%%%%%%%%%%%%%%%%%%%%%%%%
\section{Conclusion}
\label{sec:concl}
%%%%%%%%%%%%%%%%%%%%%%%%%%%%%%%%%%%%%%%%%
In this work, we implemented SPLATT, the highest performing parallel sparse tensor decomposition implementation, using Chapel.
%SPLATT is a full application for tensor decomposition, rather than a benchmark or proxy application, and is the highest performing tensor decomposition approach.
We identified features missing from Chapel that would have benefited our porting effort, such as a mutex library, varying length sub-arrays within an array of arrays, and more support for functions as first class constructs.
Through our performance study, we identified performance bottlenecks in our initial Chapel code, which involved not only Chapel's language features but also its tasking layer. We highlighted three key performance limiters:
\begin{itemize}
% I think we should just highlight the array slicing issue, as it was the most significant and affected more pieces of the code.
\item Array slicing overhead: We identified that the excessive use of array slicing was significantly degrading our performance in several parts of the code. We mitigated this by obtaining pointers to the arrays and then using pointer-arithmetic to access the required slices. This improved performance by as much as 18x but resulted in less readable code.
%\item Array creation/slicing overhead: Frequent allocation of a small array led to significant loss in performance. Furthermore, we identified that the excessive use of array slicing was significantly degrading our performance in several parts of the code. We mitigated this by obtaining pointers to the arrays and then using pointer-arithmetic to access the required slices. This improved performance by as much as 18x but resulted in less readable code.
\item Sync vs atomic variables: The use of \texttt{sync} variables under the Qthreads tasking layer for our mutex pool implementation resulted in poor parallel performance and scalability due to our very short critical sections. We redesigned the mutex pool to use \texttt{atomic} variables and achieved a 14.5x speed-up.
\item Conflicts between Qthreads and OpenMP: Using the default Qthread settings drastically degraded the performance of LAPACK routines in our code when leveraging OpenMP threads. While adjusting Qthread's affinity and spin-waiting settings improved the performance, the runtimes were still 2.3x slower than the C code, at best. Furthermore, adjusting Qthread's affinity had negative affects on other Qthreads-based routines in our code.
\end{itemize}
Our Chapel code now achieves up to 96\% of the performance of the C language implementation with respect to the critical routine, MTTKRP.

For future work, we intend to extend our Chapel port to include some of the features left out, such as support for tensors of arbitrary order and tensor mode tiling.
%We also would like to further improve some of the other routines that were not investigated in this work, such as matrix multiplication, possibly leveraging Chapel-specific profilers~\cite{dataCentric}.
We also plan to incorporate SPLATT's novel distributed-memory features~\cite{smith2016medium} for tensor decomposition in our code, leveraging Chapel's multi-locales.

%\clearpage
\bibliographystyle{IEEEtran}
\bibliography{CHIUW_Refs}

\end{document}